\documentclass[letterpaper]{article} 
\usepackage{aaai2026}  
\usepackage{times}  
\usepackage{helvet}  
\usepackage{courier}  
\usepackage[hyphens]{url}  
\usepackage{graphicx} 
\usepackage{natbib}  
\usepackage{caption} 
\usepackage{algorithm}

\usepackage{algpseudocode}
\usepackage{multirow}
\usepackage{vcell}
\usepackage{rotating}
\usepackage{booktabs}
\usepackage{xspace}
\usepackage{makecell}
\usepackage{amssymb}
\usepackage{amsmath}
\usepackage{array}
\usepackage{longtable}
\usepackage{xcolor} 

\newcommand{\sysname}{SAE\xspace}

\usepackage{newfloat}
\usepackage{listings}
\DeclareCaptionStyle{ruled}{labelfont=normalfont,labelsep=colon,strut=off} 
\floatstyle{ruled}
\newfloat{listing}{tb}{lst}{}
\floatname{listing}{Listing}
\title{Improving Sustainability of Adversarial Examples in Class-Incremental Learning}
\author{
    Taifeng Liu,
    Xinjing Liu\thanks{Corresponding author.},
    Liangqiu Dong,
    Yang Liu,
    Yilong Yang,
    Zhuo Ma$^*$
}
\affiliations{
    School of Cyber Engineering, Xidian University, China\\
    tfliu@gmx.com, 
    liuxinjing\_j@163.com, liangqiu.dong@stu.xidian.edu.cn,\\ 
    bcds2018@foxmail.com, 
    yilongyang@xidian.edu.cn, 
    mazhuo@mail.xidian.edu.cn
}

\begin{document}

\maketitle

\begin{abstract}
Current adversarial examples (AEs) are typically designed for static models.
However, with the wide application of Class-Incremental Learning (CIL), models are no longer static and need to be updated with new data distributed and labeled differently from the old ones.
As a result, existing AEs often fail after CIL updates due to significant domain drift.
In this paper, we propose \sysname to enhance the sustainability of AEs against CIL.
The core idea of \sysname is to enhance the robustness of AE semantics against domain drift by making them more similar to the target class while distinguishing them from all other classes.
Achieving this is challenging, as relying solely on the initial CIL model to optimize AE semantics often leads to overfitting.
To resolve the problem, we propose a Semantic Correction Module.
This module encourages the AE semantics to be generalized, based on a visual-language model capable of producing universal semantics.
Additionally, it incorporates the CIL model to correct the optimization direction of the AE semantics, guiding them closer to the target class.
To further reduce fluctuations in AE semantics, we propose a Filtering-and-Augmentation Module, which first identifies non-target examples with target-class semantics in the latent space and then augments them to foster more stable semantics.
Comprehensive experiments demonstrate that \sysname outperforms baselines by an average of $31.28\%$ when updated with a 9-fold increase in the number of classes.
\end{abstract}

\begin{links}
    \link{Code}{https://github.com/Jupiterliu/SAE}
\end{links}

\section{Introduction}
\label{sec:intro}

\begin{figure}[t] 
    \centering  
    \includegraphics[scale=0.41]{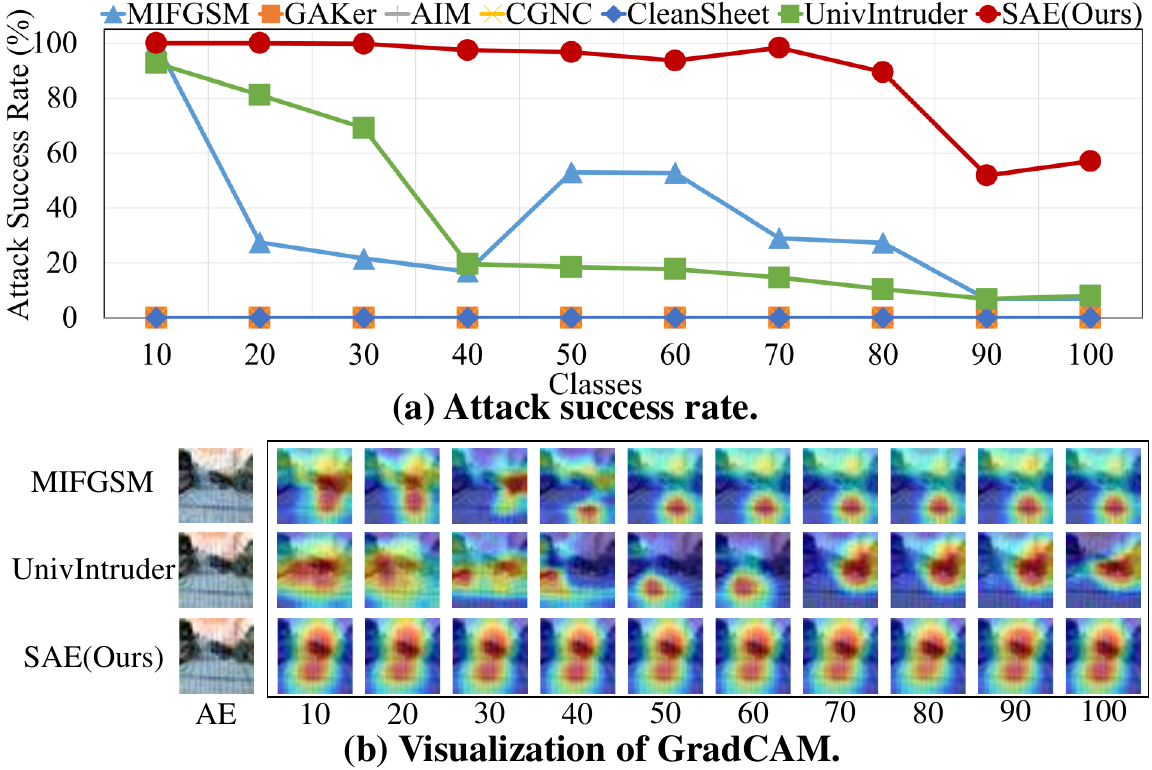}
    \caption{Attack success rate and GradCAM of different targeted adversarial attacks against CIL. The X-axis denotes the number of learned classes in CIL, with the model architecture being ResNet-32 on CIFAR-100.}
    \label{fig:attack_comparison}  
\end{figure}

Adversarial examples (AEs) pose a significant threat to machine learning models, especially in safety-critical applications like autonomous driving, healthcare, etc.~\cite{badjie2024adversarial}.
These attacks work by perturbing input data to deceive models into making incorrect predictions.
Current AEs are typically designed for static models~\cite{pelekis2025adversarial}.
However, with the advancement of Class-Incremental Learning (CIL)~\cite{zhou2024CILSurvey, zhou2024continual, zhang2025few}, models are no longer static, but sequentially updated with examples distributed and labeled differently from the old ones.
The dynamic nature of CIL causes AEs generated on the old model to become ineffective after CIL updates.
As demonstrated in Figure~\ref{fig:attack_comparison}, a small update with just $30$ classes on a ResNet-32 can lead to a significant reduction in attack success rate when evaluated with state-of-the-art (SOTA) AEs~\cite{dong2018mifgsm, li2025aim, xu2025univintruder}.
Given the broad applicability of CIL, ensuring the \textit{sustainability} of AEs in CIL scenarios is crucial, prompting the need to explore why AEs fail in CIL.

\noindent
\textbf{Domain Drift Causes AE Failures.}
Targeted AEs essentially add perturbations to input data, driving examples from the source-class domain across the model's decision boundary into the target-class domain~\cite{dong2018mifgsm, li2025aim}~\footnote{Untargeted AEs are not considered in this work, as sufficient perturbation can effectively mislead CIL models.}.
However, current AEs are typically designed for static models. 
When a CIL model is updated with new-class examples, the domains of all previous learned classes undergo significant drift~\cite{masana2022class, li2025adaptive}.
This domain drift alters both the direction and magnitude of the perturbations required to shift inputs toward the target domain. 
As a result, most AEs either mislead inputs into incorrect classes or deteriorate into benign noise, rendering them ineffective.
Although advanced adversarial attacks attempt to improve transferability by embedding semantic information of the target class~\cite{li2025aim,fang2024cgnc, sun2024gaker, ge2024cleansheet, xu2025univintruder}, they still struggle to overcome the significant domain drift affecting old classes.
The green line in Figure~\ref{fig:attack_comparison}(a) illustrates a typical semantic-based AE, which maintains an attack success rate above $20\%$ for no more than $30$ incremental classes.
Figure~\ref{fig:attack_comparison}(b) shows GradCAM~\cite{selvaraju2017gradcam} visualizations of various AEs in the CIL setting. 
As the number of learned classes increases, the effectiveness of current attacks gradually diminishes, with the regions contributing to the target class shrinking over time.

In this paper, we improve the \underline{s}ustainability of \underline{a}dversarial \underline{e}xamples in CIL by proposing \sysname. The core idea of \sysname is to enhance the robustness of AE semantics against domain drift by making them similar to the target class while distinguishing them from all other classes.
However, realizing this idea faces two main challenges:
1) AEs tend to overfit if perturbations are optimized based solely on the gradients of the initial CIL model;
2) Non-target examples unintentionally contain target-class semantics, leading to fluctuations in AE semantics.
For example, images labeled as `bicycle' may also contain `roads' when `roads' is the target class.
To address these issues, we design two modules: the Semantic Correction Module and the Filtering-and-Augmentation Module.
The first module encourages generalized AE semantics by incorporating a visual-language model that provides universal target-class semantics as an `anchor'.
Additionally, it uses the gradients of the CIL model to guide the optimization of AE semantics, ensuring consistency with the target class throughout the CIL process.
The second module detects examples with confusing semantics by calculating the cosine similarity between the non-target class and target-class examples in the latent space.
The remaining examples with low similarities are further augmented to promote more stable and generalized semantics.

We highlight our contributions as follows:
\begin{itemize}\setlength{\leftmargin}{0pt}
    \item To the best of our knowledge, we are the first to investigate the sustainability of adversarial attacks under the setting of class-incremental learning.
    \item We propose \sysname to improve the sustainability of AEs in CIL, which enhances AE robustness by making their semantics similar to the target class while distinguishing them from all other classes.
    \item To prevent AE semantics from overfitting, we propose a Semantic Correction Module that promotes generalized AE semantics towards the target class using a visual-language model and corrects the optimization direction based on the CIL model.
    \item We propose a Filtering-and-Augmentation Module that removes examples with confusing semantics in the latent space and augments the remaining examples to ensure stable and generalized AE semantics.
    \item Extensive experiments show that \sysname significantly outperforms existing approaches in terms of sustainability, improving the average attack success rate by $31.28\%$ after CIL with a $9$-fold increase in the number of classes.
\end{itemize}

\section{Related Works}
\label{sec:related}

\begin{figure*}[t] 
    \centering  
    \includegraphics[scale=0.491]{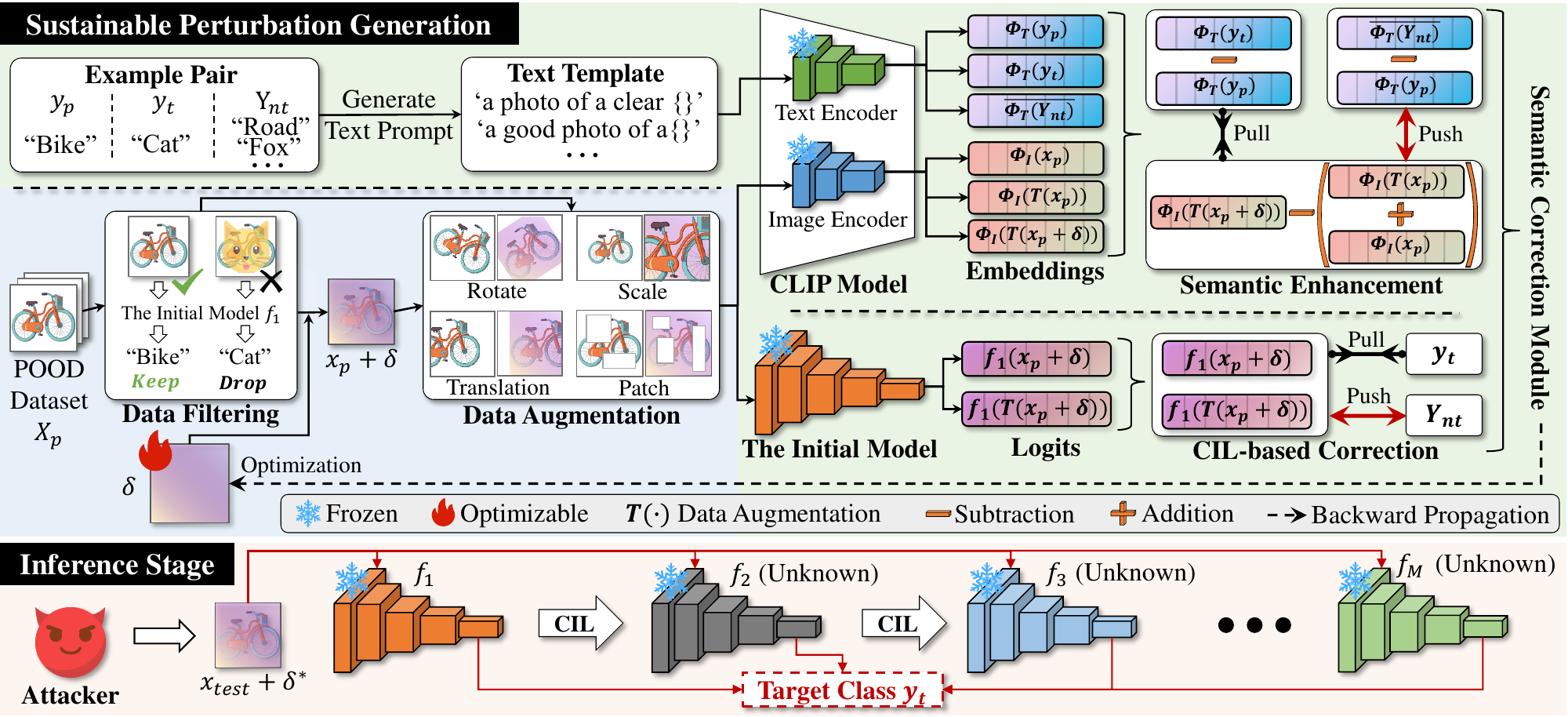}
    \caption{The overview of \sysname. }
    \label{fig:framework}  
\end{figure*}

\subsection{Class-Incremental Learning}
Class-Incremental Learning is a specific form of incremental learning (also referred to as continuous learning~\cite{de2022continuouslearning} or lifelong learning~\cite{chen2018lifelonglearning})~\cite{wang2024cl_survey}, where the model learns to classify new classes incrementally.
CIL allows users to update the model once new-class data is available. 
This makes CIL more adaptable and suitable for dynamic environments~\cite{zhou2024class, ashtekar2025class}, making it a prominent and extensively studied paradigm in incremental learning.

Current researches of CIL mainly focus on reducing catastrophic forgetting on old-class data.
There are mainly five types of CIL methods: 
1) Finetune is a typical baseline in CIL, which only utilizes the cross-entropy loss in new tasks to update the model while ignoring former tasks. It suffers from severe forgetting of former tasks;
2) Replay-based methods, such as Replay~\cite{ratcliff1990replay} and RMM~\cite{liu2021rmm}, which store interleave old-class data with new data; 
3) Knowledge distillation leverages soft targets from previous models as regularization, such as iCaRL~\cite{rebuffi2017icarl}, PodNet~\cite{douillard2020podnet}, and LKD~\cite{Gao2025LKD}; 
4) Dynamic network, such as DER~\cite{buzzega2020der}, Foster~\cite{wang2022foster}, MEMO~\cite{zhou2022memo},  L2P~\cite{wang2022l2p}, and TagFex~\cite{Zheng2025TagFex}, which dynamically expand the model architecture or adjust parameters to accommodate new tasks; 
5) Model rectification aims to reduce the biased prediction of incremental learners, such as BiC~\cite{wu2019bic}, WA~\cite{zhao2020wa}, and MoAL~\cite{gao2025MoAL}.

\subsection{Targeted Transferable Adversarial Attack}
Targeted transferable adversarial attacks represent a more challenging scenario where the adversary aims to mislead models into specific incorrect classes.
Existing targeted attacks can be broadly categorized into \emph{Iterative Attacks} and \emph{Generative Attacks}~\cite{badjie2024adversarial, li2025aim}.
Iterative attacks craft AEs based on logit-oriented loss functions, often leveraging gradients and surrogate models to enhance transferability.
Traditional targeted AEs like fast gradient sign method (FGSM)~\cite{goodfellow2014fgsm} and the momentum iterative FGSM (MIFGSM)~\cite{dong2018mifgsm} exploit the model's gradients to maximize the loss.
Recent iterative attacks focus more on the transferability of attacks to models with different architectures by extracting more robust features, such as CFM~\cite{byun2023cfm}, CleanSheet~\cite{ge2024cleansheet}, UnivIntruder~\cite{xu2025univintruder}, and LTT~\cite{weng2025LTT}.
Generative attacks have gained popularity recently due to better transferability, which produces AEs by generative models, such as TTP~\cite{naseer2021TTP}, TTAA~\cite{wang2023ttaa,sun2023dllttaa}, CGNC~\cite{fang2024cgnc}, GAKer~\cite{sun2024gaker}, AdvDiffVLM~\cite{guo2024advdiffvlm}, and AIM~\cite{li2025aim}, which leverage abundant semantic information from the latent feature space.
Although recent AEs have demonstrated strong transferability across models, they have not considered the impact of catastrophic forgetting in CIL.

\section{Approach}
\label{sec:approach}

\subsection{Problem Formulation}
\subsubsection{Class-Incremental Learning.}
In CIL, a model is trained sequentially on $M$ tasks, with each task introducing a set of new classes.
Let $f_i(\cdot)$ represent the model trained on the $i$-th task, where $i\in {1,2,...,M}$. 
The model is trained incrementally such that it learns the new classes in each task while maintaining its ability to classify all previously learned classes.

\subsubsection{Adversary's Capability \& Goals.}
The adversary can access the initial CIL model $f_1(\cdot)$ and the complete set of labels $\{y_t, Y_{nt}\}$ in the CIL process.
$y_t$ represents the target class, and $Y_{nt}$ includes all other classes learned in CIL. 
Importantly, the adversary has no access to the CIL training set, nor any information regarding the training process used to update the model.
To perform an attack, the adversary who knows CIL label can easily access public and label-mismatched data, i.e., public out-of-distribution (POOD) datasets, without knowing the distribution of the training data.
Moreover, the adversary has access to a public pre-trained visual-language model for semantic extraction.
Given a test dataset $X_{test}$ and any model $f_i(\cdot)$ within the CIL process, the adversary aims to generate a universal perturbation $\delta$: 
For any image $x\in X_{test}$, the CIL model $f_i(\cdot)$ misclassifies into a specific target class $y_t$, when applying $\delta$ on $x$.
The optimization objective is formally expressed as:
\begin{equation}
\label{eq:objective}
\begin{matrix}
    \delta^* = \mathop{\arg\min}\limits_{\delta} \mathbb{E}_{x \in X_{text}} [\mathcal{L}(f_i(x + \delta), y_t)], \forall i \ge 1 \\
    \text{s.t.} \quad ||\delta|| \le \epsilon
\end{matrix}
\end{equation}
where $\mathcal{L}(\cdot)$ is the loss function.
$\delta$ is constrained by an $l_\infty$-norm perturbation budget $\epsilon$ to ensure stealthy.

\subsection{Overview of \sysname}
Based on the core idea of semantic generalization, we design the \sysname framework, as illustrated in Figure~\ref{fig:framework}.
On the above is sustainable perturbation generation, which consists of two main modules: 1) Filtering-and-Augmentation (blue area) to address semantic fluctuation by detecting examples with confusing semantics; 2) Semantics Correction Module (green area) to optimize adversarial perturbations, ensuring sustainable target-class semantics.
At the inference stage (pink area), the adversarial perturbation, once optimized, is applied to any updated CIL model that provides black-box access, with the goal of misclassifying the perturbed examples into the target class $y_t$.
For clarity, in the following, we first introduce the optimization of AEs, followed by data filtering and augmentation.

\subsection{Semantic Correction Module}
\label{sec:strategy_1}
This part aims to generate universal semantics to optimize AEs, based on the CIL model.
CLIP, a pre-trained visual-language model, is widely considered to be capable of extracting and generating stable semantic representations for specific classes~\cite{radford2021openclip}~\footnote{CLIP is trained on billions of image-text pairs, making it particularly well-suited for zero-shot classification.}.
Previous attacks~\cite{xu2025univintruder, fang2024cgnc} also demonstrate that CLIP can generate stable targeted semantics.
Therefore, we apply CLIP to provide universal semantic information, which aligns images and text in a shared embedding space.

\begin{algorithm}[t]
    \caption{Pseudocode of Perturbation Optimization} \label{alg:optimization}
    \begin{algorithmic}[1]
        \Require initial CIL model $f_1(\cdot)$, CLIP encoders $\{\Phi_T(\cdot), \Phi_I(\cdot)\}$, POOD dataset $D_{p}$, target class $y_t$, non-target classes $Y_{nt}$, constraint $\epsilon$, Filtering Function $F_\text{filter}(\cdot)$, Augmentation Function $F_{\text{aug}}(\cdot)$.
        \Ensure Optimized perturbation $\delta$.
        \State Initialize $\delta \sim$ Gaussian(0, 1);
        \For{ each example $x_p, y_p \in D_p$}
            \While {$F_\text{filter}(x_p) > \sigma$}
                \State $\hat{x}_p = F_{aug}(x_p)$;
                \State $x'_p = F_{aug}(x_p + \delta)$;
                \State $D_{t} = \Phi_T(y_t) - \Phi_T(y_p)$;
                \State $D_{nt} = \overline{\Phi_T(Y_{nt})} - \Phi_T(y_p)$;
                \State $D_{adv} = \Phi_I(x'_p) - \Phi_I(x_p) - \Phi_I(\hat{x}_p)$;
                \State $\text{sim}_{pos}=\frac{D_{adv} \cdot D_{t}}{||D_{adv}|| \ ||D_{t}||}$;
                \State $\text{sim}_{neg}=\frac{D_{adv} \cdot D_{nt}}{||D_{adv}|| \ ||D_{nt}||}$;
                \State $\mathcal{L} = \mathcal{L}_\text{CLIP} + \mathcal{L}_\text{Surr}$; \# Referring Eq.~\ref{eq:clip_loss} and Eq.~\ref{eq:surr_loss}
                \State $\delta \gets \delta-\alpha \bigtriangledown_{\delta}\mathcal{L}$;
                \State $\delta \gets$ clamp($\delta, \epsilon$);
            \EndWhile
        \EndFor
    \end{algorithmic}
\end{algorithm}

\subsubsection{CLIP-Based Semantic Enhancement.}
To provide abundant semantics of non-target class and make target-class semantics distinguishable, we access a POOD dataset, denoted as $D_{p} = \{X_p, Y_p\}$.
For any $y_p \in Y_p$, $y_p \neq y_t$ and $y_p \notin Y_{nt}$.
In CLIP, the mapping of images and texts is realized by two encoders, i.e., the text encoder $\Phi_T(\cdot)$ and the image encoder $\Phi_I(\cdot)$, which accept image or text input and obtain embeddings.
Inspired by \cite{xu2025univintruder}, we calculate the target ($D_t$), non-target ($D_{nt}$), and adversarial ($D_{adv}$) semantic directions based on the embeddings of the image-text pairs extracted by corresponding encoders (Line 6-8 in Algorithm~\ref{alg:optimization}).
Then, \sysname calculates two types of similarity: 1) The positive similarity between the perturbed examples $(x_p+\delta)$ and the target-class text $y_t$ (pull); 2) The negative similarity between the perturbed examples and the non-target class texts $Y_{nt}$ (push)~\footnote{For notational simplicity, only here we use the notation of $y_p$, $y_t$ and $Y_{nt}$ to denote the texts of the corresponding classes.}.
Notably, for each $y_p$, the similarities need to be computed individually, and the optimization should iterate $|Y_p|$ times. Figure~\ref{fig:framework} illustrates optimization for a single $y_p$.
Based on the similarities, we define $\mathcal{L}_\text{CLIP}$ for optimizing $\delta$:
\begin{equation}
\label{eq:clip_loss}
\begin{matrix}
    \mathcal{L}_\text{CLIP} = -log(\frac{\text{sim}_{pos}}{\text{sim}_{pos} + \text{sim}_{neg}})
\end{matrix}
\end{equation}
That is, $\mathcal{L}_\text{CLIP}$ guides the AE to align with the target class's semantic embeddings while ensuring they are distant from the semantics of non-target classes.
We detail the computation of $\text{sim}_{pos}$, $\text{sim}_{neg}$ in Line 6-10 of Algorithm~\ref{alg:optimization}.
$\overline{\Phi_T(Y_{nt})}$ represents the average of all the embeddings of non-target classes, and its size is the same as that of $\Phi_T(y_{p})$.

\subsubsection{CIL-Based Correction.}
Only the static semantics provided by CLIP are hard to sustain due to the semantic drift.
Thus, we further refine them using gradients from the initial model.
It comes from the fact that guidance from the initial model remains effective due to the preserved gradient through distillation or orthogonal projection in CIL~\cite{zhou2024CILSurvey}.
Based on $f_1(\cdot)$, we obtain logits of all perturbed examples and then compute the Binary Cross-Entropy (BCE) loss between the logits and the target class $y_t$:
\begin{equation}
\label{eq:surr_loss}
\begin{matrix}
    \mathcal{L}_\text{Surr} = -log(p_{y_t})-\sum_{y_{nt}\in Y_{nt}} log(1-p_{y_{nt}})
\end{matrix}
\end{equation}
where $p_{y_t}$ and $p_{Y_{nt}}$ represent the predicted probability of the target class and non-target classes made by $f_1(\cdot)$, respectively.
$\mathcal{L}_\text{Surr}$ ensures that the adversarial perturbations generated based on the surrogate model not only mislead the model prediction but also help improve the sustainability of the attack across evolving models in the CIL process.

\subsection{Filtering-and-Augmentation Module}
To reference confusing semantics, we first gather several typical examples of class $y_t$ to obtain embeddings that represent the target class.
These examples can be collected from the POOD datasets, denoted as $X_c$.
Each example in the dataset $X_p$ is first filtered with reference to $X_c$ and then augmented before being fed into the CLIP encoder $\Phi_I(x)$ and the surrogate model $f_1(\cdot)$.
Thus, we compute the embeddings for $X_c$ and $X_p$ based on $f_1(\cdot)$, respectively: $E_c\gets \frac{1}{|X_c|}\sum f_1(X_c)$, $E_p\gets f_1(X_p)$, where the embeddings are extracted from the second-to-last layer, $E_c$ is the averaged embeddings.
Then, for each $x_i \in X_p$, compute cosine similarity between its embedding and $E_c$:
\begin{equation}
    F_{\text{filter}}(x_i) = \frac{E_p^i\cdot E_c}{\left \|E_p^i \right \| \cdot \left \| E_c\right\|}
\end{equation}
where $E_p^i$ is the embedding of $x_i$.
The output of $F_{\text{filter}}(x_i)$ is further normalized to the range $(0, 1)$.
$x_i$ with similarity $\text{sim}(x_i, E_c)$ larger than $\sigma$ are considered to have semantics of the target class, which are filtered.
$\sigma$ is determined based on the statistics of cosine similarity.
Line 3 in Algorithm~\ref{alg:optimization} indicates that only examples that lack confusing semantics for the target class can be further applied to optimize the adversarial perturbation.

We then apply random augmentations, including rotation, scaling, translation, and patching, to increase variations in the features of examples, as inspired by prior works~\cite{xu2025univintruder}.
These augmentations are applied randomly with varying strength and parameters to enhance the diversity of examples, thereby reducing overfitting of semantics of AEs.
As illustrated in Line 4-5 in Algorithm~\ref{alg:optimization}, the remained example $x_p$ and perturbed example $x_p + \delta$ are all augmented.

\section{Experiments}
\label{sec:results}

\begin{table*}
\centering
\footnotesize
\newcommand{\rot}[1]{\rotatebox{90}{#1}}
\newcommand{\multirot}[1]{\multirow{8}{*}{\rot{#1}}}
\caption{Average SASR across ten target classes for various attacks and CIL methods. All CIL methods are trained using the CIFAR-100 and ImageNet-100 datasets. Clean Acc denotes the average accuracy of models trained across ten tasks.}
\label{tab:all_ASR_CIFAR100}
\setlength{\tabcolsep}{2.54mm}{
\begin{tabular}{cccccccccccc} 
\toprule
\multicolumn{1}{c}{\textbf{Dataset}}    & \textbf{\textbf{Attack}}                & \textbf{Finetune}     & \textbf{Replay}        & \textbf{MEMO~}         & \textbf{DER}           & \textbf{Foster}        & \textbf{WA}            & \textbf{BiC}           & \textbf{iCaRl}         & \textbf{PodNet}        & \textbf{AVG.}           \\ 
\hline\hline
\multirot{CIFAR-100}    & Clean Acc                               & 26.23                 & 58.38                  & 71.69                  & 70.90                  & 64.89                  & 65.54                  & 60.39                  & 62.85                  & 55.65                  & 59.61                   \\ 
\cline{2-12}
                                        & MIFGSM                                  & 7.29                  & 13.41                  & 21.71                  & 36.85                  & 17.93                  & 16.51                  & 18.95                  & 17.94                  & 44.17                  & 21.64                   \\
                                        & GAKer                                   & 0.04                  & 0.02                   & 0.01                   & 0.00                   & 0.00                   & 0.04                   & 0.00                   & 0.00                   & 5.43                   & 0.62                    \\
                                        & AIM                                     & \underline{\textbf{9.94}} & 30.96                  & 0.05                   & 0.04                   & 48.47                  & 56.20                  & 0.04                   & 28.15                  & 0.02                   & 19.32                   \\
                                        & CGNC                                    & 0.57                  & 0.12                   & 0.00                   & 0.00                   & 0.48                   & 0.30                   & 0.00                   & 0.00                   & 0.08                   & 0.17                    \\
                                        & CleanSheet                              & 1.23                  & 2.31                   & 3.80                   & 1.16                   & 2.02                   & 2.71                   & 4.02                   & 6.00                   & 7.88                   & 3.46                    \\
                                        & UnivIntruder                            & 4.85                  & 31.23                  & 34.82                  & 40.23                  & 31.23                  & 41.82                  & 33.62                  & 32.43                  & 42.20                  & 32.49                   \\ 
\cline{2-12}
                                        & \textbf{\sysname (Ours)} & \textbf{9.38}         & \textbf{\underline{37.56}} & \textbf{\underline{52.72}} & \textbf{\underline{63.07}} & \textbf{\underline{49.47}} & \textbf{\underline{63.51}} & \textbf{\underline{53.14}} & \textbf{\underline{36.49}} & \textbf{\underline{85.30}} & \textbf{\underline{50.07}}  \\ 
\midrule
\multirot{ImageNet-100} & Clean Acc                               & 24.91                 & 58.75                  & 71.69                  & 74.14                  & 68.40                  & 67.39                  & 61.58                  & 57.89                  & 67.28                  & 61.34                   \\ 
\cline{2-12}
                                        & MIFGSM                                  & \textbf{\underline{9.80}} & 9.36                   & 16.05                  & 26.23                  & 22.46                  & 15.05                  & 9.48                   & 11.08                  & 68.02                  & 20.84                   \\
                                        & GAKer                                   & 0.14                  & 1.08                   & 0.33                   & 0.17                   & 0.22                   & 0.00                   & 0.00                   & 0.00                   & 0.17                   & 0.23                    \\
                                        & AIM                                     & 0.23                  & 4.46                   & 2.26                   & 4.34                   & 1.77                   & 1.09                   & 1.60                   & 1.60                   & 3.60                   & 2.33                    \\
                                        & CGNC                                    & 0.07                  & 0.11                   & 0.05                   & 0.06                   & 0.05                   & 0.05                   & 0.06                   & 0.01                   & 0.06                   & 0.06                    \\
                                        & CleanSheet                              & 0.00                  & 0.00                   & 0.00                   & 0.00                   & 0.00                   & 0.00                   & 0.00                   & 0.47                   & 0.00                   & 0.05                    \\
                                        & UnivIntruder                            & 0.34                  & 4.17                   & 1.63                   & 3.36                   & 2.82                   & 2.44                   & 0.75                   & 0.93                   & 2.53                   & 2.11                    \\ 
\cline{2-12}
                                        & \textbf{\sysname (Ours)} & \textbf{6.04}         & \textbf{\underline{18.92}} & \textbf{\underline{32.88}} & \textbf{\underline{32.62}} & \textbf{\underline{22.53}} & \textbf{\underline{24.09}} & \textbf{\underline{24.70}} & \textbf{\underline{21.74}} & \textbf{\underline{83.89}} & \textbf{\underline{29.71}}  \\
\bottomrule
\end{tabular}}
\end{table*}

\subsection{Experimental Setup}
The setups of datasets, models, CIL methods and baselines are introduced, followed by evaluation metrics and baselines compared with \sysname.

\subsubsection{Datasets and Models.}
We consider two widely used benchmark datasets: CIFAR-100~\cite{cifar100} and ImageNet-100~\cite{icarl2017Imagenet100}, each comprising 100 classes. 
Following recent CIL research~\cite{zhou2024CILSurvey}, both datasets are partitioned into 10 groups of 10 classes to simulate the CIL process. 
The target class is selected from the first 10 classes. 
We use ResNet-32 for CIFAR-100 and ResNet-50 for ImageNet-100, due to their effectiveness in image classification and suitability in CIL.

Specifically, we use Tiny-ImageNet~\cite{le2015tinyimagenet}, which contains 200 classes, as the POOD dataset for CIFAR-100.
For ImageNet-100, we use ImageNet-1K, which includes 1000 classes. 
To ensure a clean separation, we exclude classes from ImageNet-1K that overlap with ImageNet-100, guaranteeing no overlap in classes or images between the POOD dataset and the CIL training set.

\subsubsection{CIL Methods and Attack Baselines.}
We evaluate \sysname on nine representative CIL methods, categorized into five paradigms: 
1) Finetune, usually used as the baseline of CIL~\cite{zhou2024CILSurvey};
2) replay-based methods, including Replay~\cite{ratcliff1990replay};
3) dynamic network-based methods, such as MEMO~\cite{zhou2022memo}, DER~\cite{buzzega2020der}, and Foster~\cite{wang2022foster}; 
4) model rectification-based methods, including WA~\cite{zhao2020wa} and BiC~\cite{wu2019bic}; 
5) knowledge distillation-based methods, represented by iCaRL~\cite{rebuffi2017icarl} and PodNet~\cite{douillard2020podnet}. 
These methods are implemented using a public CIL benchmark framework\footnote{https://github.com/LAMDA-CL/PyCIL}.

To ensure fair comparison, we select representative targeted transferable attacks from both generative and iterative paradigms as baselines. 
Specifically, we adopt AIM~\cite{li2025aim}, GAKer~\cite{sun2024gaker}, and CGNC~\cite{fang2024cgnc} as generative baseline attacks. 
For iterative baseline attacks, we include MIFGSM~\cite{dong2018mifgsm}, CleanSheet~\cite{ge2024cleansheet}, and UnivIntruder~\cite{xu2025univintruder}.

\begin{figure*}[t] 
    \centering  
    \includegraphics[scale=0.557]{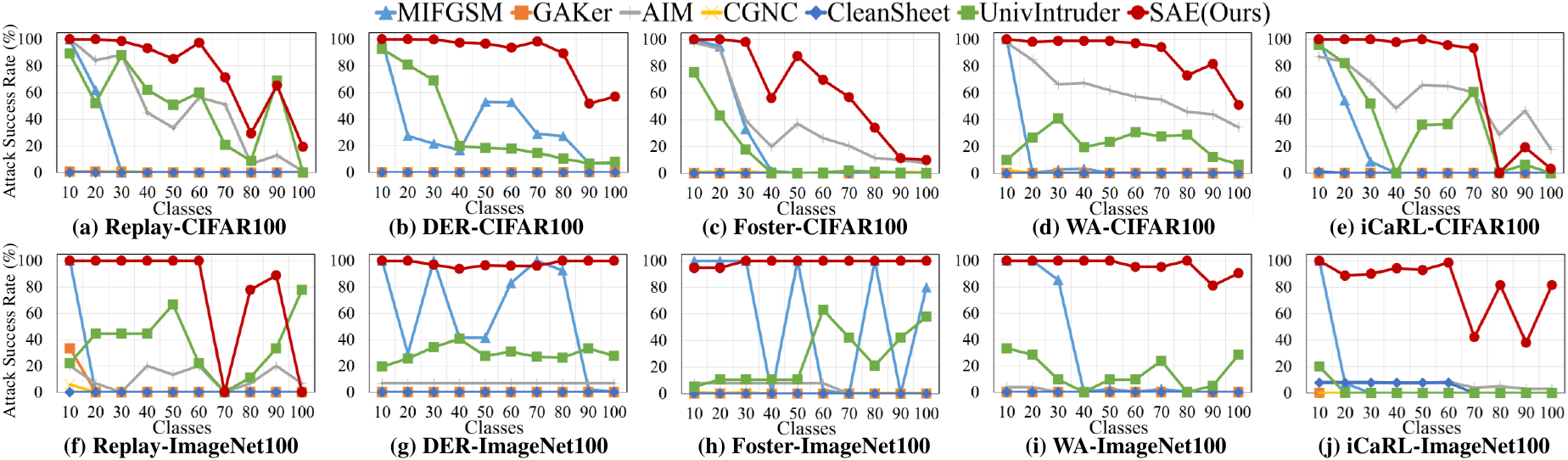} 
    \caption{ASR curves for both baseline attacks and our attack across various CIL methods. Each subfigure illustrates the ASR across incremental tasks. Subfigures (a)–(e) present results for the CIFAR-100 dataset with `skyscraper' as the target class. Subfigures (f)–(j) show the corresponding results for the ImageNet-100 dataset using `candy store' as the target class.}
    \label{fig:sustainability}  
\end{figure*}

\subsubsection{Implementation Details and Metrics.}
We adopt CLIP as implemented in OpenCLIP~\cite{radford2021openclip}. 
Specifically, we use the ViT-B-32 model pre-trained on the LAION-2B dataset~\cite{schuhmann2022laion2b}. 
For perturbation optimization, we employ the Adam optimizer~\cite{diederik2014adam} with a learning rate of $0.01$, a weight decay of $1 \times 10^{-5}$, and a batch size of $256$. 
The optimization is conducted for $50$ epochs.
Additionally, constraint for epsilon $\epsilon$ is set to $32/255$, following the standard setting for black-box transferable attacks~\cite{chen2023practical, xu2025univintruder}. 
$\sigma$ for filtering is set $0.7$ and embeddings are extracted from the second-to-last layer of $f_1(\cdot)$.
All experiments are conducted on an RTX 4060 GPU with 8GB of memory.

We evaluate performance using two metrics: Attack Success Rate (ASR) and Sustainable Attack Success Rate (SASR). 
ASR denotes the proportion of perturbed test examples that are classified into the target class by the model updated after the $i$-th task $f_i(\cdot)$.
SASR evaluates the sustainability of adversarial examples across the entire CIL process, which is defined as:
\begin{equation}
    \text{SASR} = \frac{1}{M \times C}\sum^{M}_{i=1}\sum^{C}_{j=1}\mathbb{I}[f_i(x_j+\delta)=y_t]
\end{equation}
where $\mathbb{I}(\cdot)$ is an indicator function, which is $1$ when classified correctly and $0$ otherwise, $x_j \in X_{test}$, $C=|X_{test}|$.

\begin{table*}
\centering
\footnotesize
\caption{SASR of \sysname across different ten attack target classes and five CIL methods, evaluated using models trained on the CIFAR-100 dataset. The results demonstrate the effectiveness of \sysname in maintaining high ASR across diverse target classes.}
\label{tab:ASR_per_target_class}
\setlength{\tabcolsep}{2.1mm}{
\begin{tabular}{ccccccccccc} 
\toprule
\textbf{CIL Method~} & \textbf{`road'} & \textbf{`palm\_tree'} & \textbf{`snake'} & \textbf{`bicycle'} & \textbf{`cloud'} & \textbf{`table'} & \textbf{`train'} & \textbf{`rabbit'} & \textbf{`shrew'} & \textbf{`skyscraper'}  \\ 
\hline\hline
Replay               & 27.17           & 36.58                 & 41.17            & 80.89              & 27.82            & 24.18            & 36.24            & 15.40             & 10.11            & 76.00                  \\
DER                  & 62.92           & 96.53                 & 24.96            & 99.57              & 5.41             & 88.79            & 94.18            & 19.70             & 50.19            & 88.39                  \\
Foster               & 23.04           & 80.13                 & 14.04            & 72.64              & 17.63            & 48.30            & 99.82            & 46.96             & 29.84            & 62.35                  \\
WA                   & 43.67           & 99.46                 & 79.73            & 99.98              & 36.76            & 67.27            & 57.58            & 35.51             & 26.02            & 89.16                  \\
iCaRL                & 28.77           & 13.52                 & 36.10            & 82.94              & 3.47             & 31.68            & 61.48            & 22.57             & 13.45            & 70.97                  \\ 
\hline
\textbf{AVG.}        & \textbf{37.12}  & \textbf{65.24}        & \textbf{39.20}   & \textbf{87.20}     & \textbf{18.22}   & \textbf{52.04}   & \textbf{69.86}   & \textbf{28.03}    & \textbf{25.92}   & \textbf{77.37}         \\
\bottomrule
\end{tabular}}
\end{table*}

\subsection{Results}
\subsubsection{Sustainability.}
We compare \sysname with baseline attacks under various CIL methods. 
As illustrated in Table~\ref{tab:all_ASR_CIFAR100}, attack performance is evaluated using the average SASR across ten target classes on CIFAR-100 and ImageNet-100 datasets.
The best results are highlighted via underline formatting. 
\sysname consistently outperforms competing approaches under most CIL methods, achieving substantial improvements.
On CIFAR-100, our method surpasses the average of all baselines by $37.12\%$ across all CIL methods. 
Similar trends are observed on the ImageNet-100 dataset, where \sysname continues to outperform all baselines by $25.44\%$ average.
As a result, our method achieves an average performance improvement of $31.28\%$ across both datasets.
In this experiment, we observe variation in SASR across different CIL methods. Specifically, for AEs of \sysname and some baselines such as UnivIntruder and AIM, their SASRs on Finetune are relatively low.
We hypothesize that this variability is due to relatively stronger catastrophic forgetting in such a CIL method, which applies no strategies to mitigate the forgetting.
In such a case, the CIL model struggles to correctly classify most samples (Clean Acc $26.23\%$), thus rendering the performance of AEs irrelevant.

We further analyze the performance of generative and iterative AEs.
Among generative attacks, including GAKer, AIM, and CGNC, AIM demonstrates better sustainability than GAKer and CGNC due to its semantic injection module, which embeds target-class semantics into each layer of the generator for improved stability.
For iterative attacks, i.e., MIFGSM, CleanSheet, and \sysname, the results show that \sysname significantly outperforms both MIFGSM and CleanSheet. MIFGSM relies heavily on gradients of $f_1(\cdot)$, making it prone to overfitting. On the other hand, CleanSheet focuses on universal target-class semantics but neglects the CIL model. 
Overall, \sysname outperforms all the baselines, highlighting the effectiveness of our Semantic Correction Module, which prevents overfitting while ensuring the robustness of adversarial examples.

We also evaluate the performance of different adversarial attacks using the ASR across various task updates in the CIL model, as shown in Figure~\ref{fig:sustainability}. The x-axis represents the number of learned classes, while the y-axis shows the ASR for each attack. Each subplot corresponds to a CIL method.
From the results, ASR curves for \sysname are consistently higher than baselines and exhibit significantly less fluctuation, indicating that the target-class semantics in \sysname are more stable.
When excluding the initial ASR in SASR, \sysname achieves $35.45\%$ avg. SASR ($+28.76\%$ vs. all baselines).
Thus, \sysname demonstrates greater robustness against CIL. 
It can also be observed that in CIL methods such as Replay and iCaRL, \sysname experiences fluctuation in ASR when the number of learned classes exceeds about $70$.
This decline occurs because CIL methods rely on fixed-length caches to store previous data or model parameters to mitigate catastrophic forgetting~\cite{zhou2024CILSurvey}. Once the cache is full, the model's ability to retain earlier knowledge weakens, causing a significant drop in ASR.
Due to space constraints, we refer the interested reader to the extended version of our paper for all results.

\subsubsection{Robustness to Target Class.}
To assess the robustness of \sysname across different target classes, we performed targeted attacks on a range of classes. 
As shown in Table~\ref{tab:ASR_per_target_class}, we report the SASR across various CIL methods trained on the CIFAR-100 dataset. 
The results demonstrate that \sysname maintains strong robustness across diverse target classes, with the average SASR exceeding $50\%$ for all targets.
A closer analysis reveals that \sysname achieves notably high performance on most target classes, such as an average SASR of $87.2\%$ for `bicycle' and $77.37\%$ for `skyscraper'. 
For a small subset of classes, such as `shrew' and `cloud', the method yields relatively lower SASR values due to the confusing semantics of these classes.
This variation stems from CIL class degradation rather than the shortage of our attack.

\begin{figure}[t] 
    \centering  
    \includegraphics[scale=0.39]{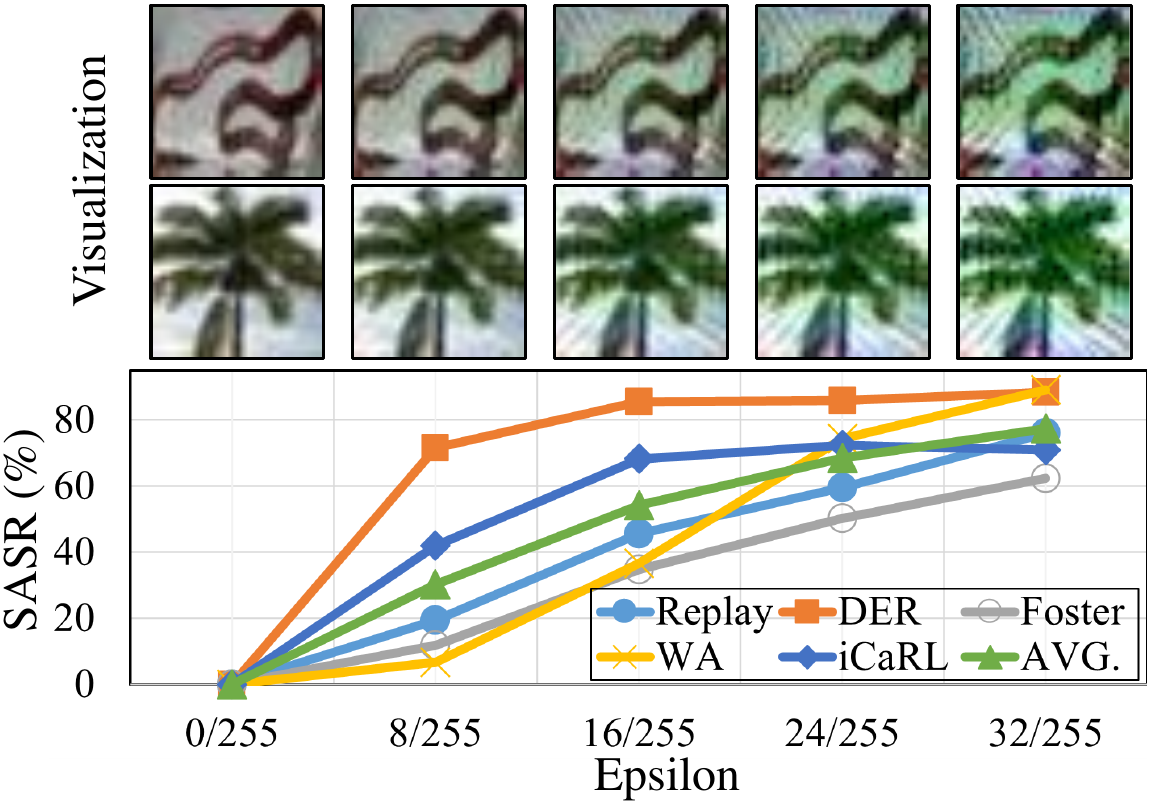}
    \caption{Perturbation’s constraints and SASR on CIFAR-100. The target class is `skyscraper'.}
    \label{fig:visibility}  
\end{figure}

\begin{table}[t]
\centering
\footnotesize
\caption{Evaluation of four adversarial defenses on CIFAR-100 under clean and perturbed inputs. The target class is `skyscraper'. The results also compare the clean accuracy (Clean) of CIL and attack performance (SASR) without and with defenses.}
\label{tab:defense}
\setlength{\tabcolsep}{2mm}{
\renewcommand\arraystretch{1}
\begin{tabular}{ccccc} 
\toprule
\multirow{2}{*}{\textbf{Defense}}                             & \multicolumn{2}{c}{\textbf{Foster}} & \multicolumn{2}{c}{\textbf{iCaRL}}  \\
                                                              & \textbf{Clean} & \textbf{SASR}      & \textbf{Clean} & \textbf{SASR}      \\ 
\hline\hline
without Defense                                               & 71.15          & 62.35              & 75.28          & 70.97              \\ 
\hline
Adversarial Training                                           & 58.30          & 51.26              & 61.57          & 41.26              \\
\begin{tabular}[c]{@{}c@{}}Input Transformations\end{tabular} & 32.78          & 38.45              & 37.95          & 52.69              \\
\textcolor[rgb]{0.2,0.2,0.2}{Feature Denoising}               & 43.10          & 58.52              & 52.33          & 64.80              \\
\textcolor[rgb]{0.2,0.2,0.2}{Random Smoothing}                & 31.25          & 42.74              & 47.05          & 60.68              \\
\bottomrule
\end{tabular}}
\end{table}

\subsubsection{Perturbation’s Visibility.}
In this paper, we constrain adversarial perturbations using the $l_\infty$-norm. 
To explore the impact of different perturbation constraints on performance, we conduct experiments under constraints of $8/255$, $16/255$, $24/255$, and $32/255$. 
The results, including the SASR and images illustrating perturbation perceptibility, are shown in Figure~\ref{fig:visibility}. 
The SASR exhibits a minor drop from $77.37\%$ to $68.48\%$ as the perturbation bound decreases from $32/255$ to $24/255$. 
It retains a strong performance of $54.15\%$ even when the $l_\infty$-norm is further reduced to $16/255$, indicating that \sysname can generate effective adversarial examples even under more restrictive and imperceptible perturbations.
Similar to CleanSheet~\cite{ge2024cleansheet} and UniverIntruder~\cite{xu2025univintruder}, the adversarial examples generated by \sysname under various constraints are visually indistinguishable from their natural counterparts and remain imperceptible to human observers.

\begin{figure}[t] 
    \centering  
    \includegraphics[scale=0.4]{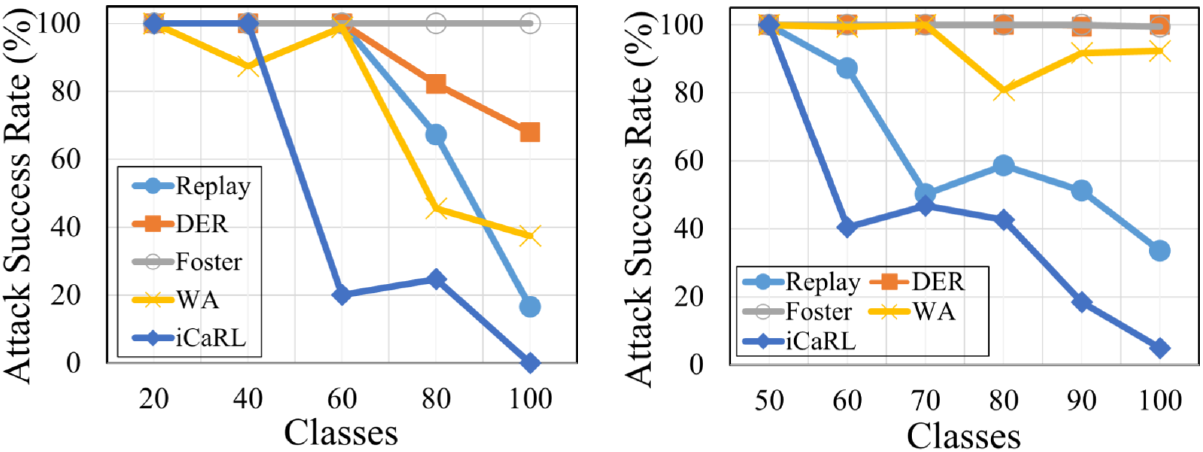}  
    \caption{ASR curves for different CIL settings on CIFAR-100. The left plot illustrates ASR across five tasks, with CIL learning 20 new classes per task. The right plot illustrates ASR across six tasks, where CIL initially learns 50 classes, followed by incremental learning of 10 new classes per task.}
    \label{fig:diff_task_abliation}  
\end{figure}

\begin{table}[t]
\centering
\footnotesize
\caption{Ablation study of proposed modules. No Module indicates that only the gradients of the initial CIL model are used to optimize the adversarial samples. We also report the affect of various $\sigma$ in FAM.}
\label{tab:ablation_module}
\setlength{\tabcolsep}{1mm}{
\renewcommand\arraystretch{1}
\begin{tabular}{cc|cc} 
\toprule
\textbf{Strategy}                    & \textbf{SASR} & \textbf{\textbf{Strategy}}                & \textbf{\textbf{SASR}}  \\ 
\hline\hline
No Module                            & 21.64         & FAM Only~($\sigma=0.7$)      & 23.13                   \\
SCM Only                             & 42.82         & SCM \& FAM ($\sigma=0.5$) & 48.91                   \\
FAM Only~($\sigma=0.5$) & 22.37         & SCM \& FAM ($\sigma=0.7$) & 50.07                   \\
\bottomrule
\end{tabular}}
\end{table}

\subsubsection{Defenses Evasion.}
In Table~\ref{tab:defense}, we select four typical defense methods that are primarily used in CIL to enhance adversarial robustness~\cite{cho2025enhancing} to evaluate \sysname, including Adversarial Training~\cite{liu2025adversarialtraining}, Image Transformations~\cite{meng2017imagetransformation},
Feature Denoising~\cite{xie2019featuredenoising},
and Random Smoothing~\cite{jeong2023smoothing}. 
The results show that \sysname maintains an average SASR of $51.30\%$.
Although these defenses enhance the robustness of models against \sysname, they introduce a trade-off between model robustness and performance. 
As a result, these defenses may only be suitable for scenarios where high model accuracy is not a critical requirement.

\subsubsection{Ablation Study.}
We examine the impact of different CIL configurations on the SASR. 
As shown in Figure~\ref{fig:diff_task_abliation}, our method maintains a consistent average SASR of $76.89\%$ across different tasks, demonstrating the robustness of our approach and its ability to maintain high performance while adapting to new classes in diverse CIL scenarios.

Under the default attack evaluation setting, we evaluate the effectiveness of the two proposed modules, as shown in Table~\ref{tab:ablation_module}. 
SCM and FAM denote the Semantic Correction Module and the Filtering-and-Augmentation Module, respectively. 
From the results, applying the Semantic Correction Module increases the SASR by $21.18\%$.
Adding the FAM further boosts \sysname, outperforming the strategy without any modules by $28.43\%$. 
We further evaluate with $\sigma=0.5$, which filters out more examples in FAM. Compared to the default setting of $\sigma=0.7$, we observe a slight decline in performance, as many samples without confusing semantics are filtered out, resulting in fewer examples available for optimization.

\section{Conclusion}
\label{sec:conclusion}
In this paper, we propose \sysname to enhance the sustainability of AEs in CIL. By addressing key challenges including overfitting and fluctuating AE semantics, we introduce two modules: the Semantic Correction Module, which uses a generative model and CIL gradients to ensure non-overfitted target-class semantics, and the Filtering-and-Augmentation Module, which eliminates examples with confusing semantics and augments the remained ones. 
Comprehensive experiments validate the effectiveness of \sysname, demonstrating improved sustainability across various CIL.

\section{Acknowledgments}
We sincerely appreciate the anonymous reviewers for their insightful comments.
This work was supported by the National Natural Science Foundation of China (U21A20464, U23A20306, U23A20307, U2436206, 62406239), the China Postdoctoral Science Foundation (No. 2023M742739), the `111 Center’ (B16037), and the Fundamental Research Funds for the Central Universities (Program No.QTZX24081).

\bibliography{aaai2026}

\onecolumn
\appendix

\section{Appendix}
\begin{figure*}[!htbp] 
    \centering  
    \includegraphics[scale=0.43]{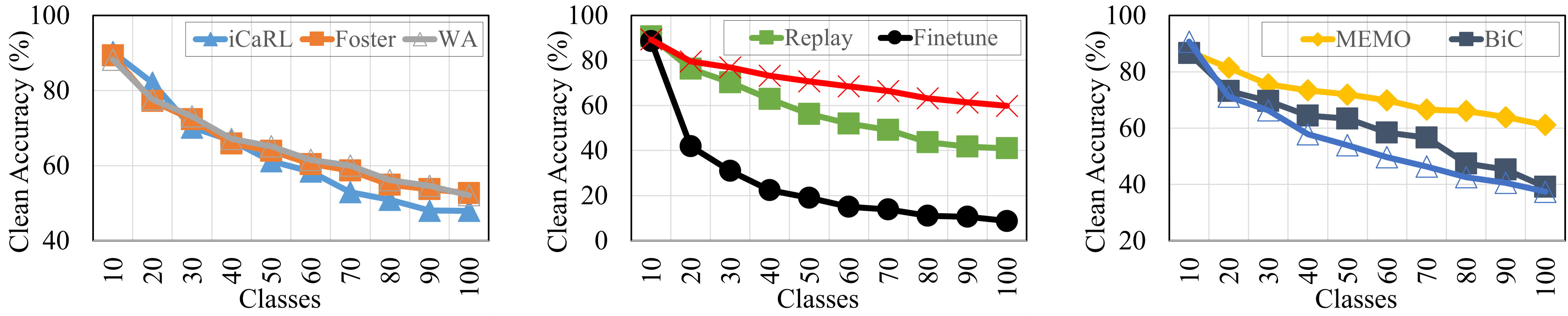} 
    \caption{Clean accuracy of nine CIL methods on the CIFAR-100 dataset.}
    \label{fig:CIL_cifar100}  
\end{figure*}

\begin{figure*}[!htbp] 
    \centering  
    \includegraphics[scale=0.43]{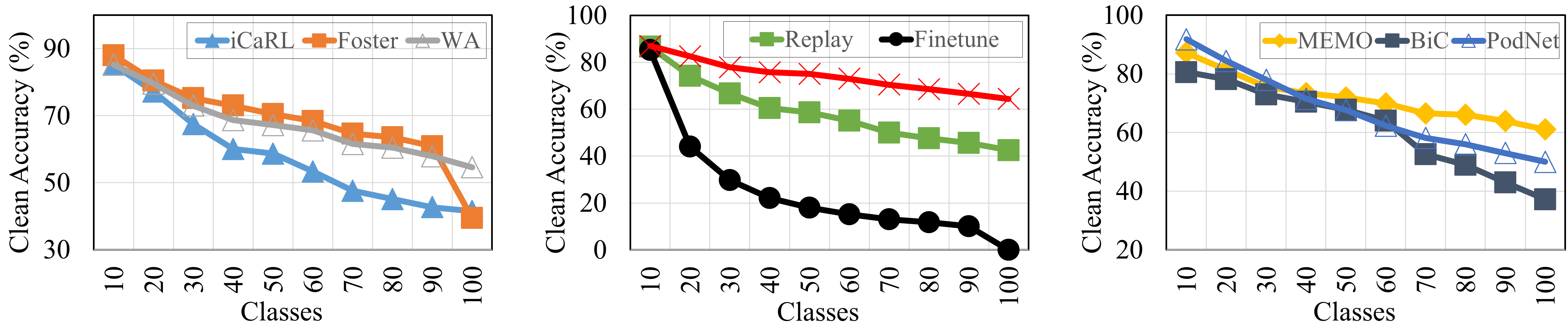} 
    \caption{Clean accuracy of nine CIL methods on the ImageNet-100 dataset.}
    \label{fig:CIL_imagenet100}  
\end{figure*}

In this section, we illustrate the clean accuracy of nine CIL methods on the CIFAR-100 (Figure~\ref{fig:CIL_cifar100}) and ImageNet-100 (Figure~\ref{fig:CIL_imagenet100}) datasets. 
Then, we present all ASR curves for both baseline attacks and the proposed attack SAE across various CIL methods. 
Each subfigure in Figure 3-20 illustrates the ASR across incremental tasks against a specific target class.
For CIFAR-100, the ten target classes are `road', `palm tree', `snake', `bicycle', `cloud', `table', `train', `rabbit', `shrew', and `skyscraper', respectively.
For ImageNet-100, the ten target classes are `mailbox', `candy store', `CRT monitor', `quail', `Greater Swiss Mountain Dog', `tennis ball', `wool', `lens cap', `ping-pong ball', and `purse', respectively.

\subsection{Performance of Class-Incremental Learning on CIFAR-100}

\subsection{Performance of Class-Incremental Learning on ImageNet-100}

\subsection{Benchmark Results of Targeted Attacks on CIFAR-100}


\begin{figure*}[!htbp] 
    \centering  
    \includegraphics[width=\textwidth,height=0.75
    \textheight,keepaspectratio]{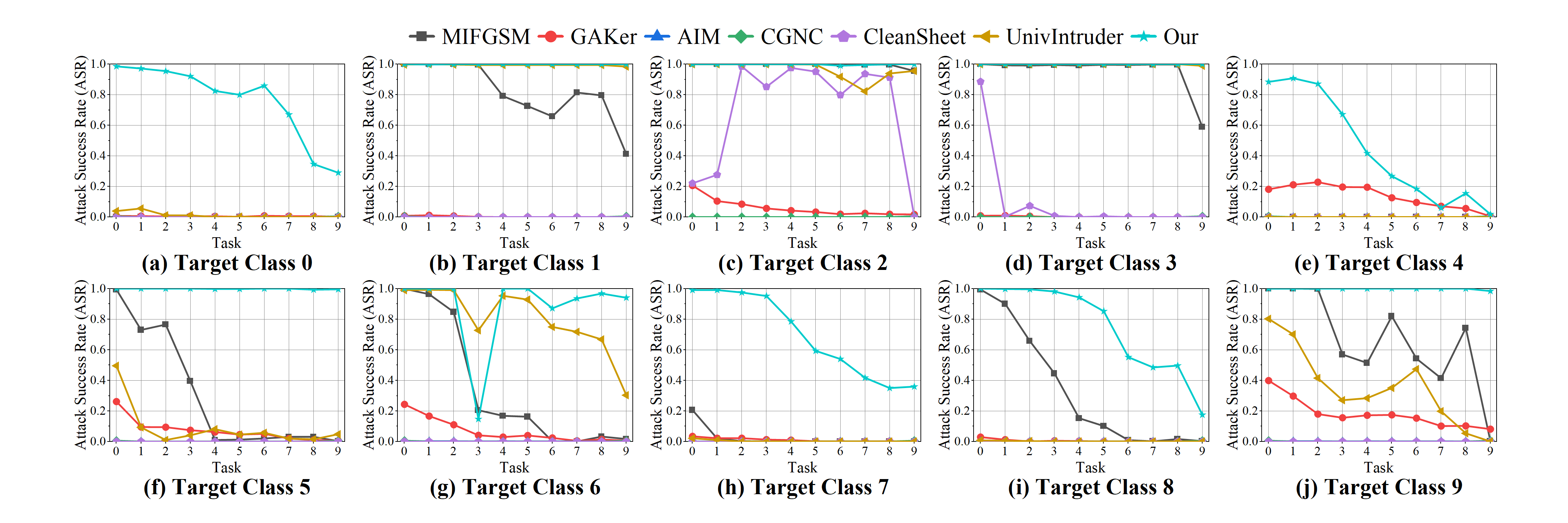} 
    \caption{Attack performance under the CIL method of \textbf{Podnet}.}
    \label{fig:cifar100_podnet}  
\end{figure*}

\begin{figure*}[!htbp] 
    \centering  
    \includegraphics[width=\textwidth,height=0.75
    \textheight,keepaspectratio]{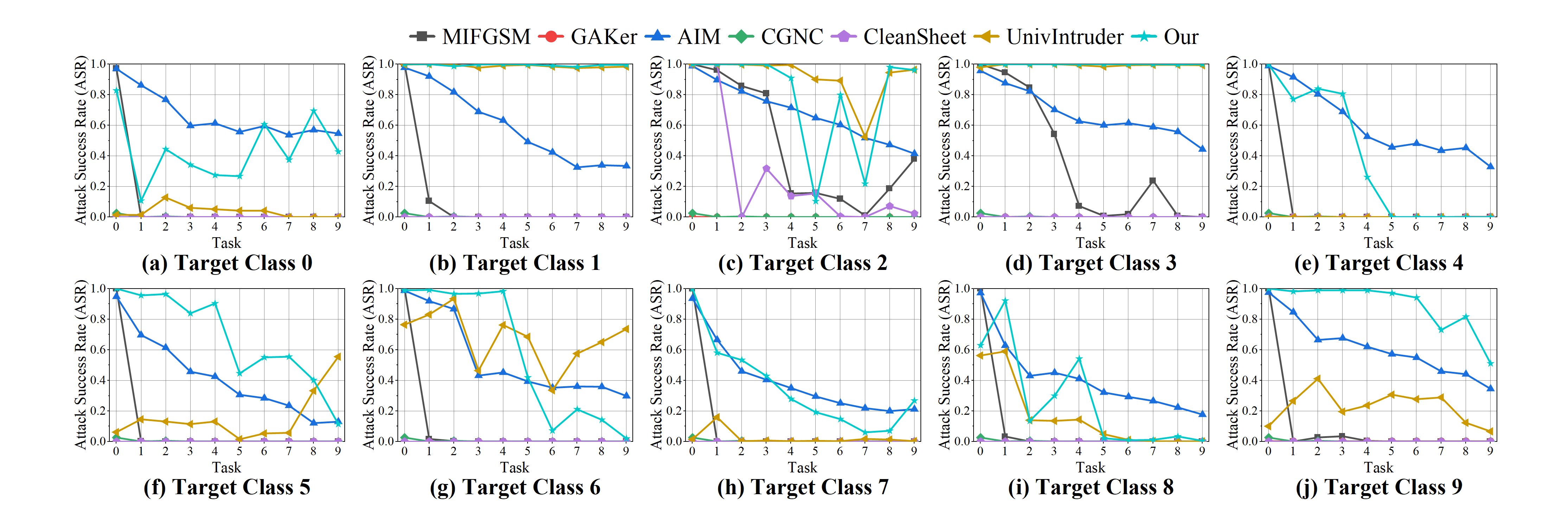} 
    \caption{Attack performance under the CIL method of \textbf{WA}.}
    \label{fig:cifar100_wa}  
\end{figure*}

\begin{figure*}[!htbp] 
    \centering  
    \includegraphics[width=\textwidth,height=0.75
    \textheight,keepaspectratio]{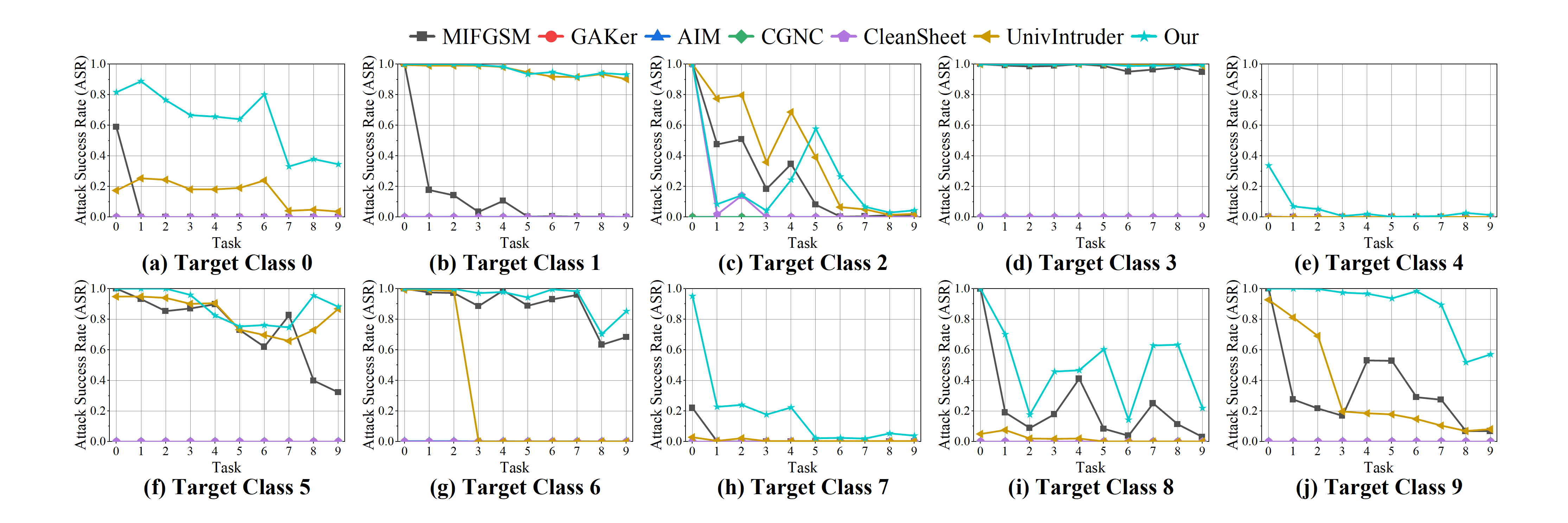} 
    \caption{Attack performance under the CIL method of \textbf{DER}.}
    \label{fig:cifar100_der}  
\end{figure*}

\begin{figure*}[!htbp] 
    \centering  
    \includegraphics[width=\textwidth,height=0.75
    \textheight,keepaspectratio]{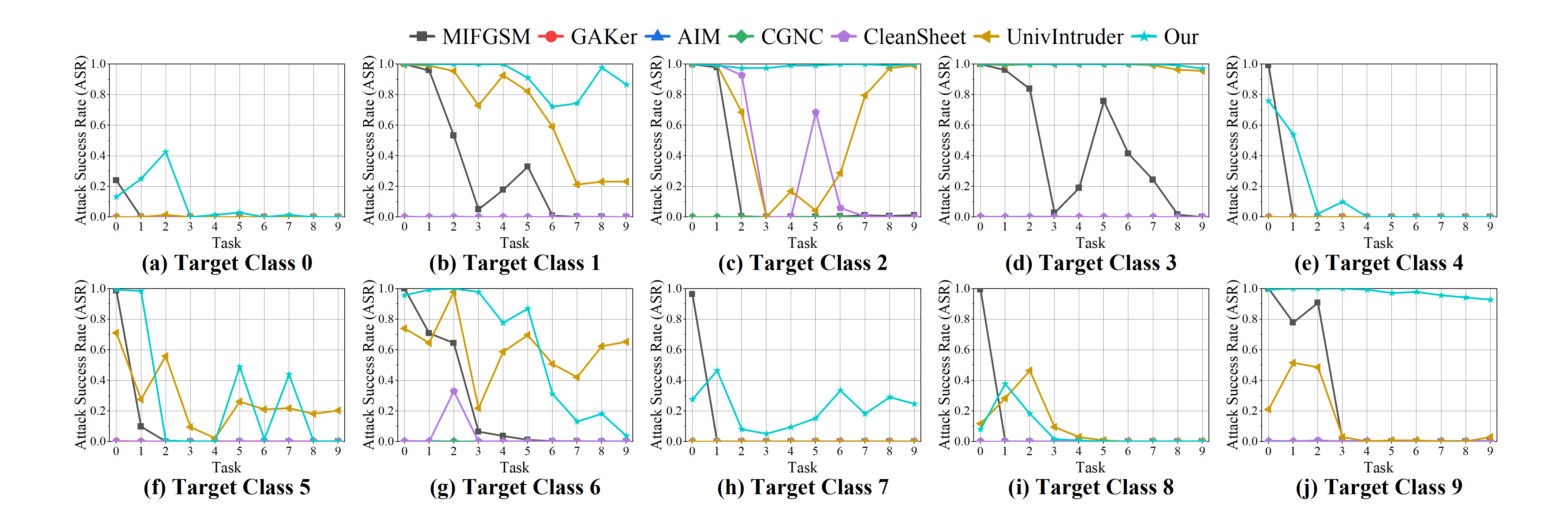} 
    \caption{Attack performance under the CIL method of \textbf{BiC}.}
    \label{fig:cifar100_bic}  
\end{figure*}

\begin{figure*}[!htbp] 
    \centering  
    \includegraphics[width=\textwidth,height=0.75
    \textheight,keepaspectratio]{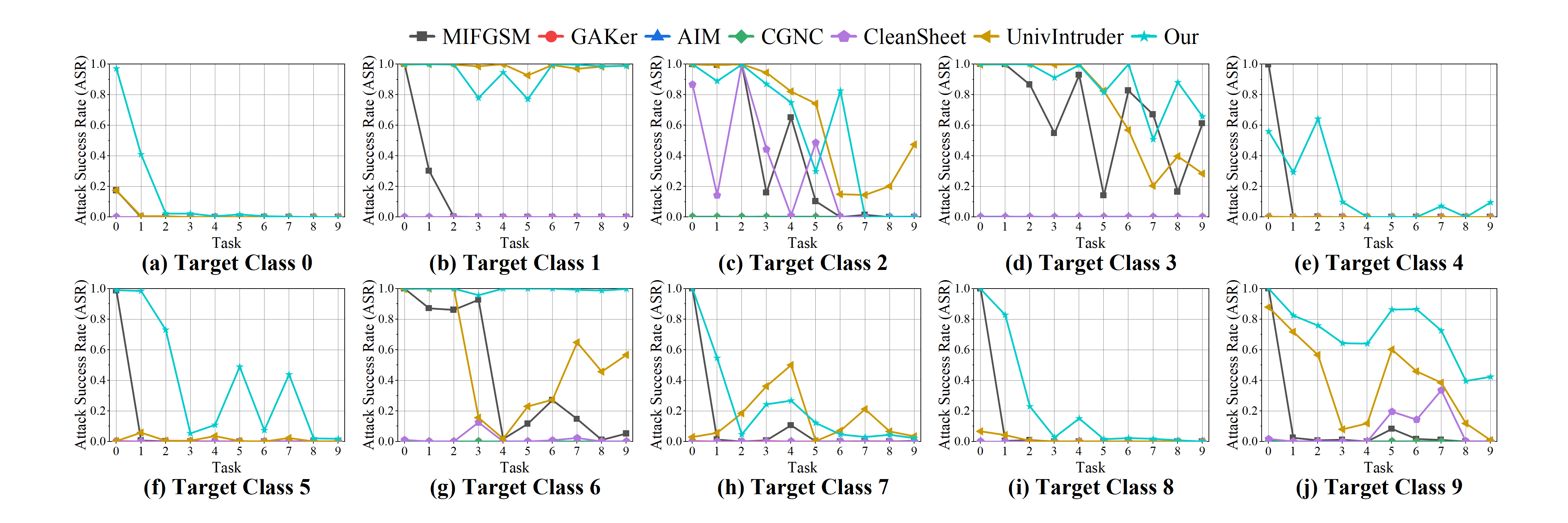} 
    \caption{Attack performance under the CIL method of \textbf{MEMO}.}
    \label{fig:cifar100_memo}  
\end{figure*}

\begin{figure*}[!htbp] 
    \centering  
    \includegraphics[width=\textwidth,height=0.75
    \textheight,keepaspectratio]{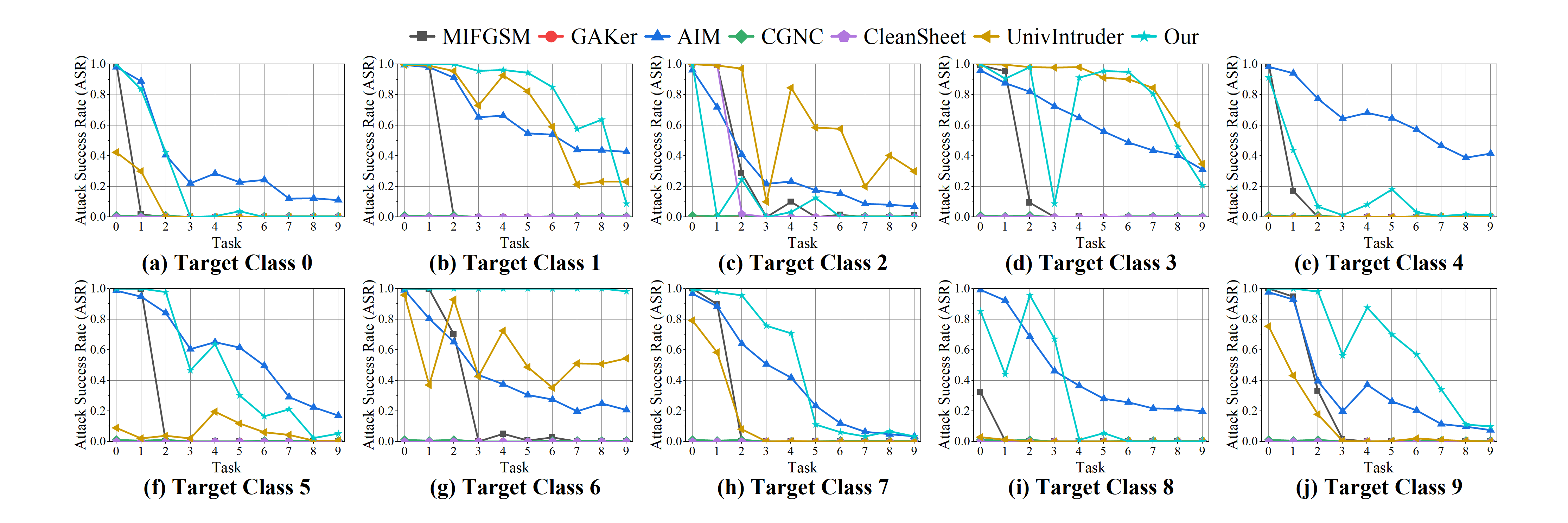} 
    \caption{Attack performance under the CIL method of \textbf{Foster}.}
    \label{fig:cifar100_foster}  
\end{figure*}

\begin{figure*}[!htbp] 
    \centering  
    \includegraphics[width=\textwidth,height=0.75
    \textheight,keepaspectratio]{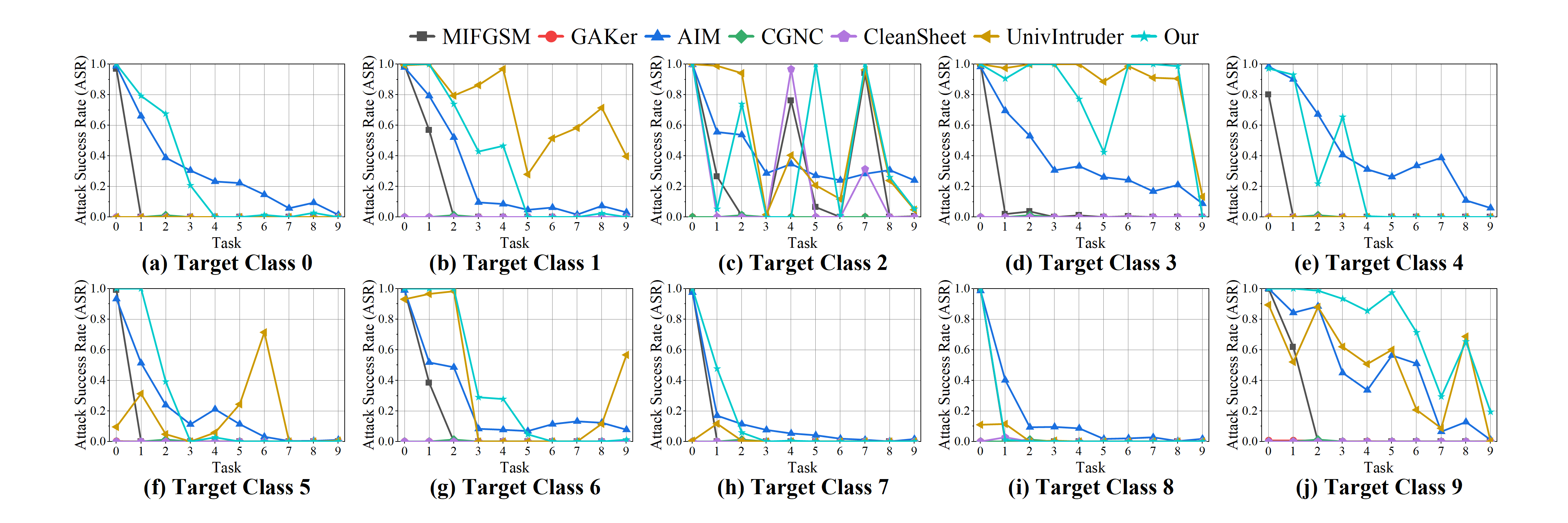} 
    \caption{Attack performance under the CIL method of \textbf{Replay}.}
    \label{fig:cifar100_replay}  
\end{figure*}

\begin{figure*}[!htbp] 
    \centering  
    \includegraphics[width=\textwidth,height=0.75
    \textheight,keepaspectratio]{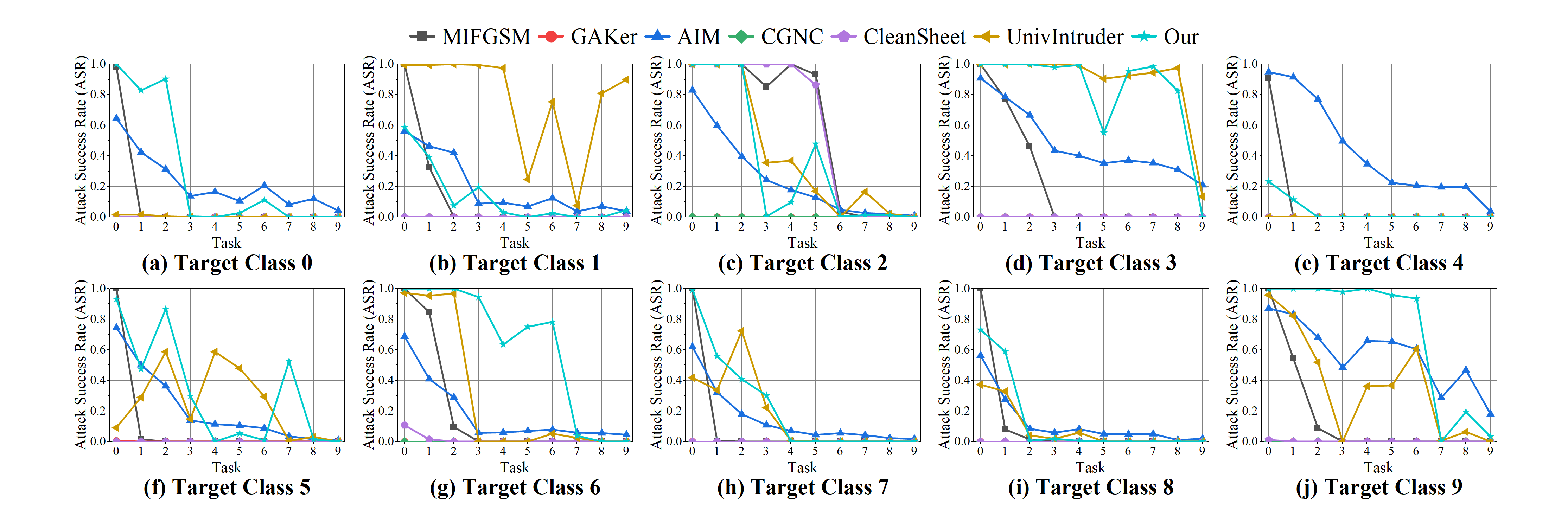} 
    \caption{Attack performance under the CIL method of \textbf{iCaRL}.}
    \label{fig:cifar100_icarl}  
\end{figure*}

\begin{figure*}[!htbp] 
    \centering  
    \includegraphics[width=\textwidth,height=0.75
    \textheight,keepaspectratio]{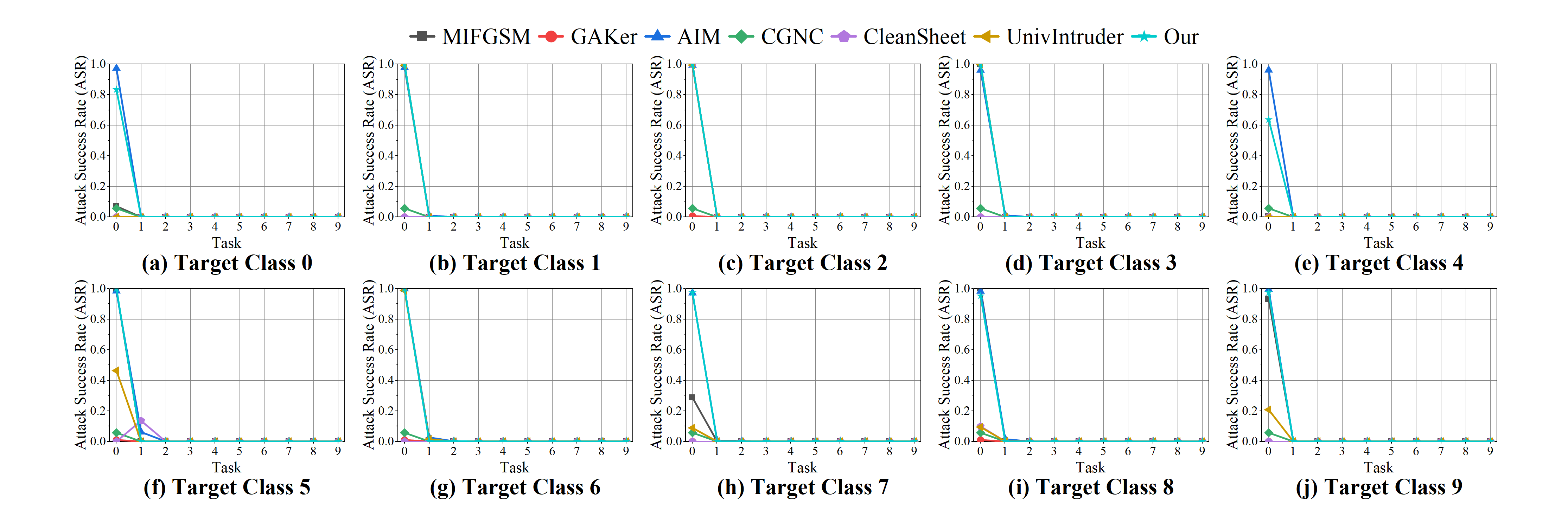} 
    \caption{Attack performance under the CIL method of \textbf{Finetune}.}
    \label{fig:cifar100_finetune}  
\end{figure*}

\clearpage

\subsection{Benchmark Results of Targeted Attacks on Imagenet-100}

\begin{figure*}[!htbp] 
    \centering  
    \includegraphics[width=\textwidth,height=0.75
    \textheight,keepaspectratio]{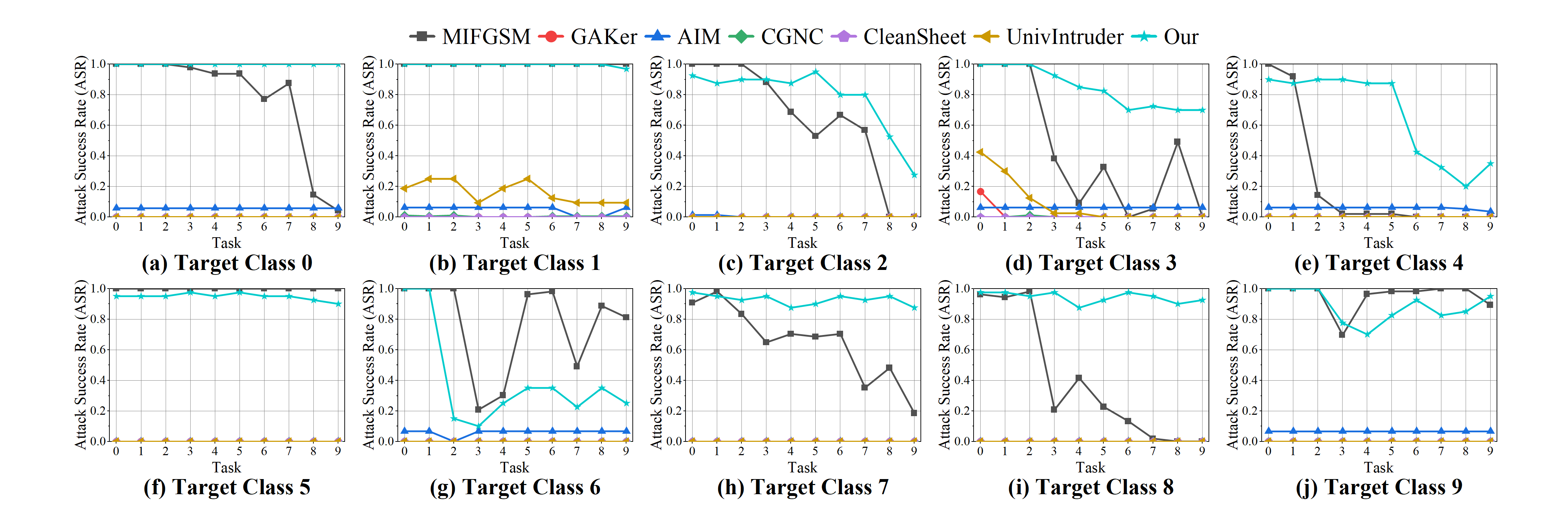} 
    \caption{Attack performance under the CIL method of \textbf{PodNet}.}
    \label{fig:imagenet100_podnet}  
\end{figure*}

\begin{figure*}[!htbp] 
    \centering  
    \includegraphics[width=\textwidth,height=0.75
    \textheight,keepaspectratio]{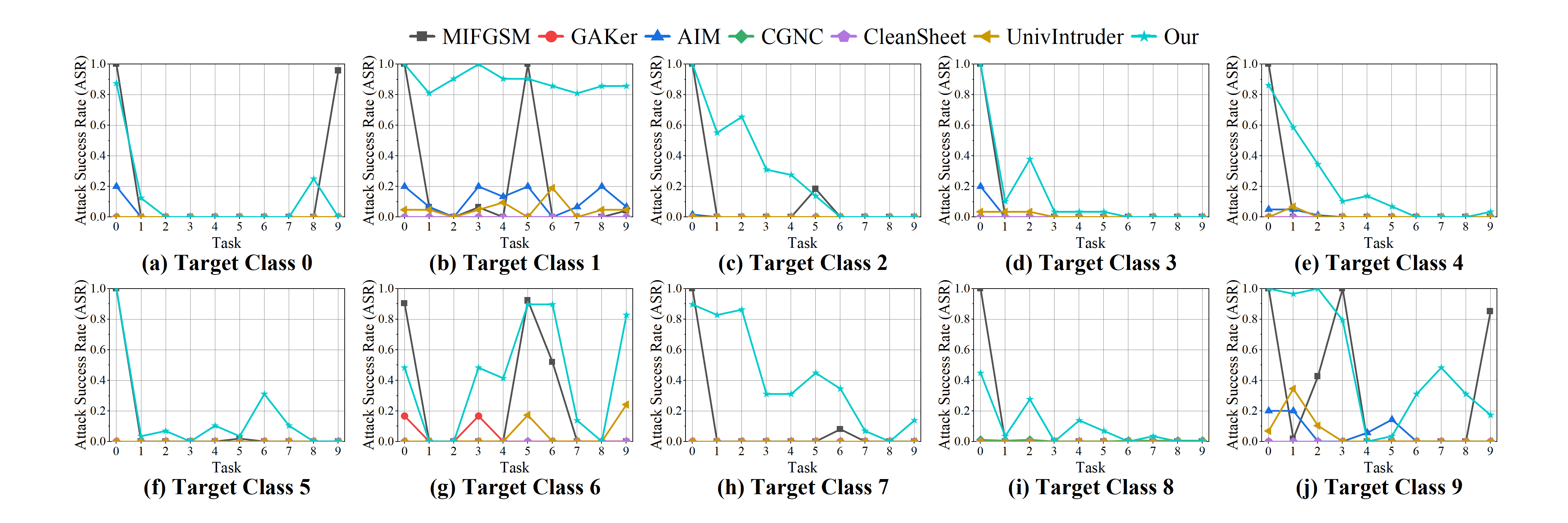} 
    \caption{Attack performance under the CIL method of \textbf{MEMO}.}
    \label{fig:imagenet100_memo}  
\end{figure*}

\begin{figure*}[!htbp] 
    \centering  
    \includegraphics[width=\textwidth,height=0.75
    \textheight,keepaspectratio]{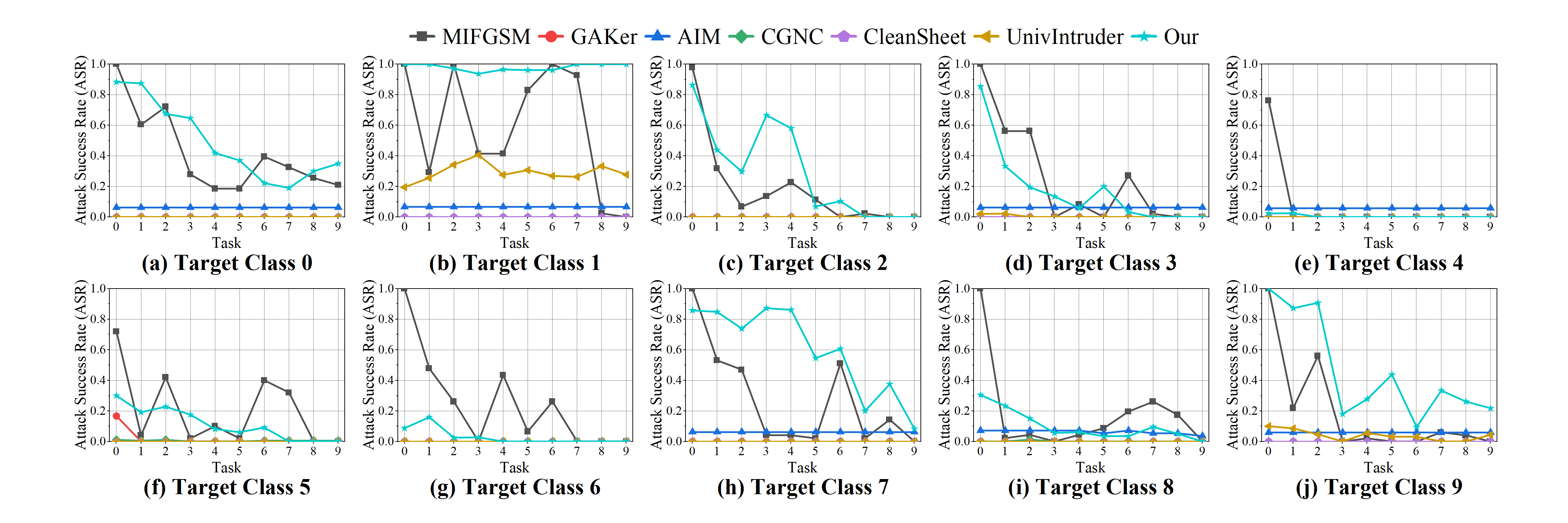} 
    \caption{Attack performance under the CIL method of \textbf{DER}.}
    \label{fig:imagenet100_der}  
\end{figure*}

\begin{figure*}[!htbp] 
    \centering  
    \includegraphics[width=\textwidth,height=0.75
    \textheight,keepaspectratio]{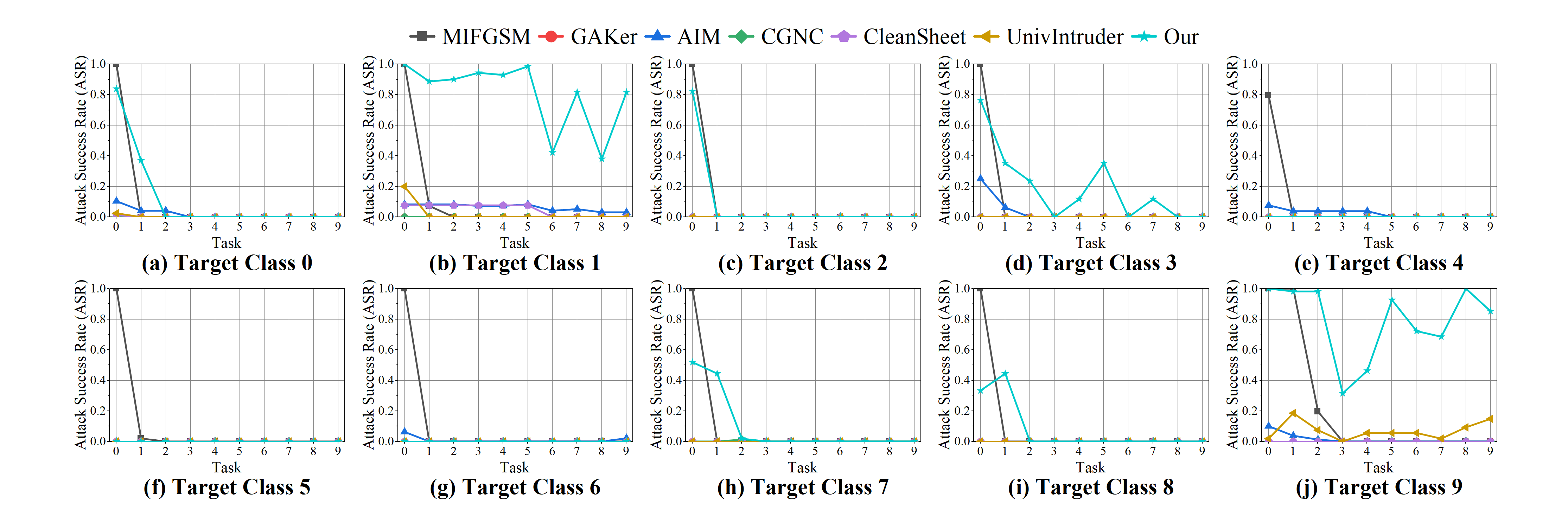} 
    \caption{Attack performance under the CIL method of \textbf{iCaRL}.}
    \label{fig:imagenet100_icarl}  
\end{figure*}

\begin{figure*}[!htbp] 
    \centering  
    \includegraphics[width=\textwidth,height=0.75
    \textheight,keepaspectratio]{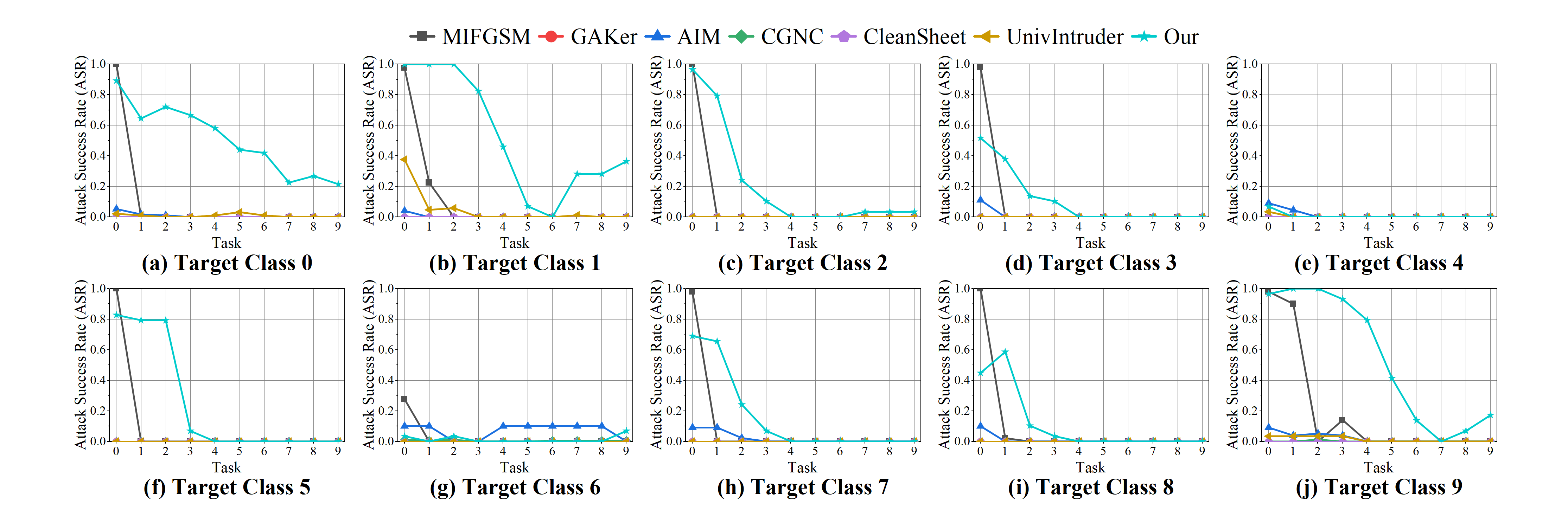} 
    \caption{Attack performance under the CIL method of \textbf{BiC}.}
    \label{fig:imagenet100_bic}  
\end{figure*}

\begin{figure*}[!htbp] 
    \centering  
    \includegraphics[width=\textwidth,height=0.75
    \textheight,keepaspectratio]{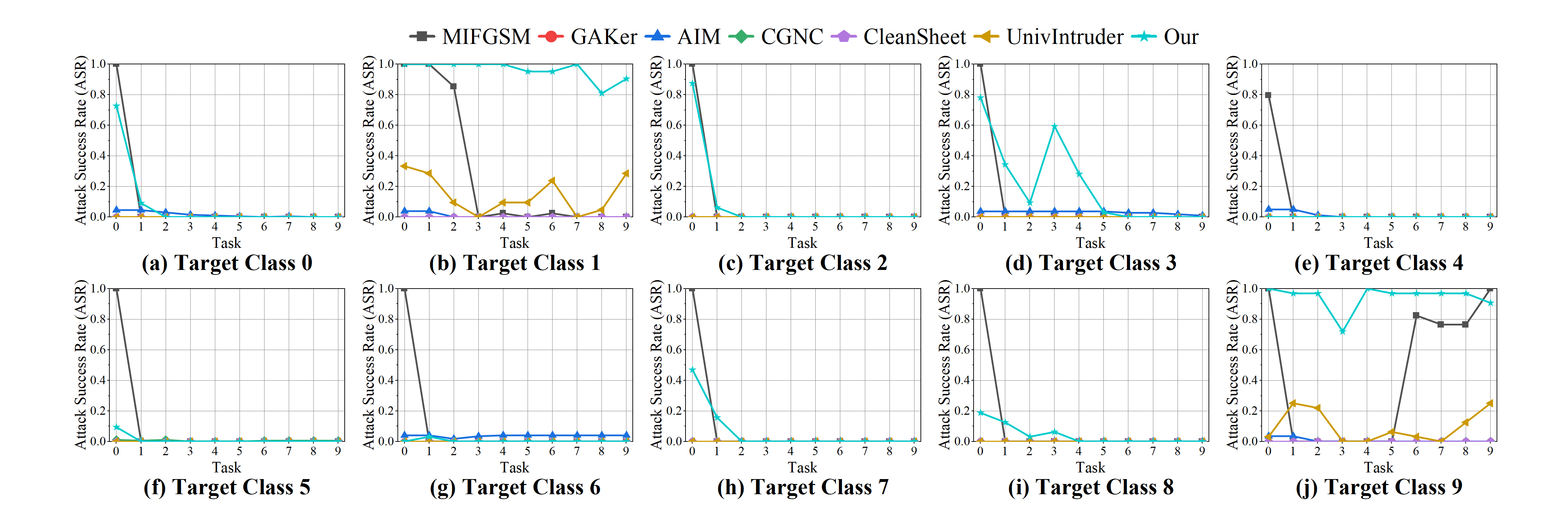} 
    \caption{Attack performance under the CIL method of \textbf{WA}.}
    \label{fig:imagenet100_wa}  
\end{figure*}

\begin{figure*}[!htbp] 
    \centering  
    \includegraphics[width=\textwidth,height=0.75
    \textheight,keepaspectratio]{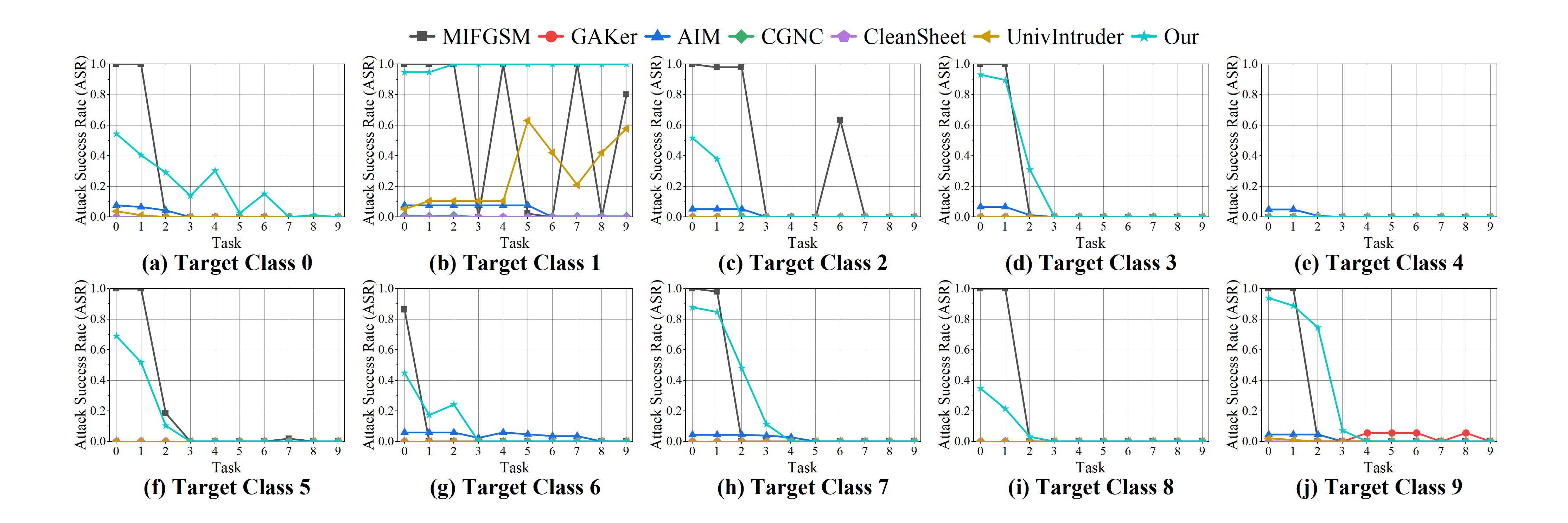} 
    \caption{Attack performance under the CIL method of \textbf{Foster}.}
    \label{fig:imagenet100_foster}  
\end{figure*}

\begin{figure*}[!htbp] 
    \centering  
    \includegraphics[width=\textwidth,height=0.75
    \textheight,keepaspectratio]{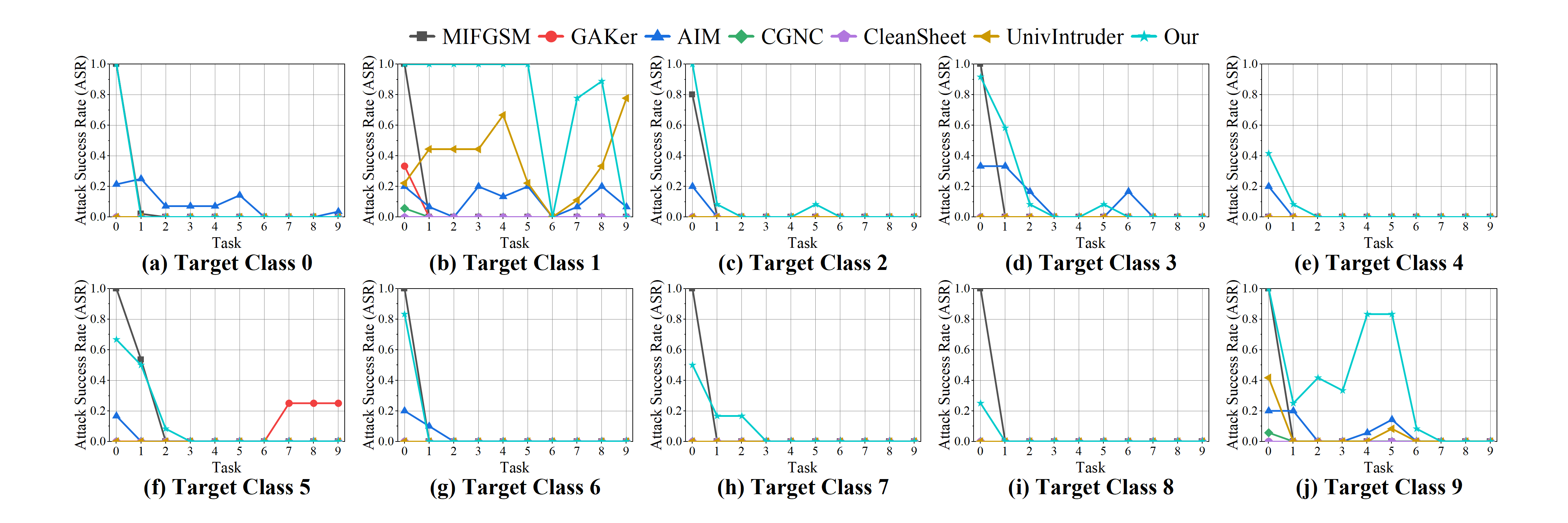} 
    \caption{Attack performance under the CIL method of \textbf{Replay}.}
    \label{fig:imagenet100_replay}  
\end{figure*}

\begin{figure*}[!htbp] 
    \centering  
    \includegraphics[width=\textwidth,height=0.75
    \textheight,keepaspectratio]{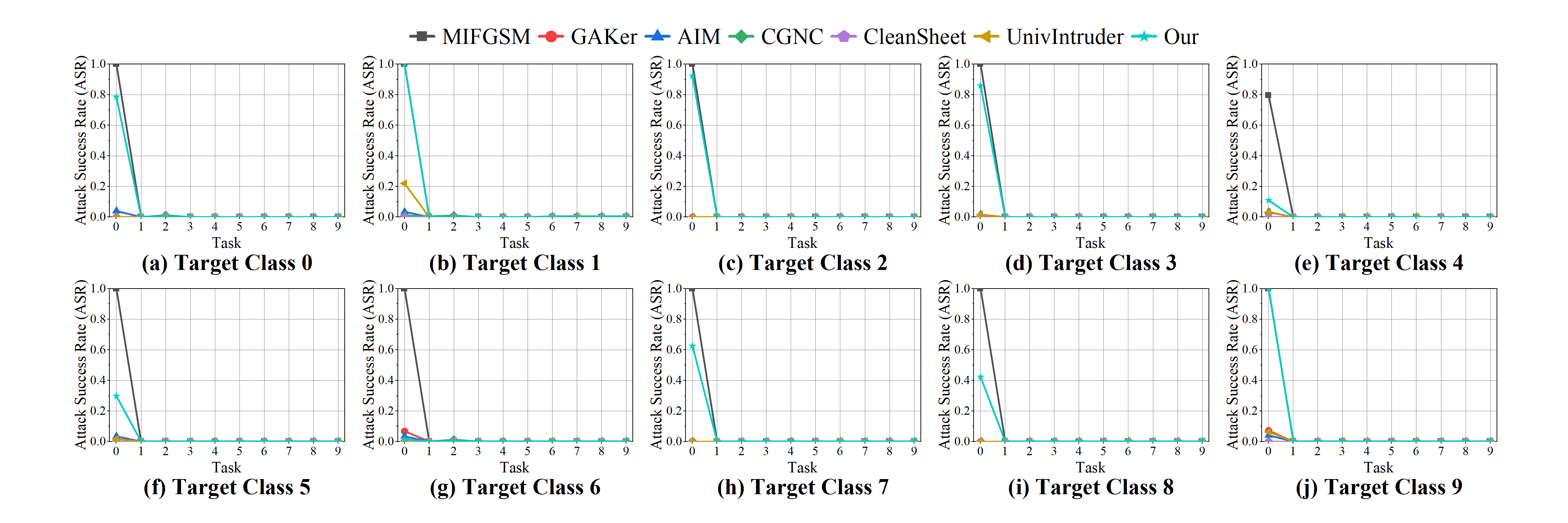} 
    \caption{Attack performance under the CIL method of \textbf{Finetune}.}
    \label{fig:imagenet100_finetune}  
\end{figure*}

\end{document}